\documentclass[aip,amsmath,amssymb,reprint]{revtex4-2}

\usepackage{graphicx}
\usepackage{dcolumn}
\usepackage{bm}

\usepackage[utf8]{inputenc}
\usepackage[T1]{fontenc}
\usepackage{mathptmx}
\usepackage{hyperref}
\usepackage{etoolbox}
\usepackage{caption}
\DeclareCaptionLabelSeparator{VertLine}{$\,\vert\,$}
\usepackage{subcaption}
\usepackage{physics}
\usepackage{siunitx}
\usepackage{comment}
\usepackage{hhline}
\usepackage{multirow}

\makeatletter
\def\@email#1#2{%
 \endgroup
 \patchcmd{\titleblock@produce}
  {\frontmatter@RRAPformat}
  {\frontmatter@RRAPformat{\produce@RRAP{*#1\href{mailto:#2}{#2}}}\frontmatter@RRAPformat}
  {}{}
}%
\renewcommand{\figurename}{\textbf{Fig.}}

\renewcommand \thefigure{\textbf{\@arabic\c@figure}}
\renewcommand \thetable{\textbf{\@arabic\c@table}}
\makeatother

\begin{document}

\preprint{AIP/123-QED}

\title{The SpinBus Architecture: Scaling Spin Qubits with Electron Shuttling}
\author{Matthias K\"unne}
\altaffiliation{These authors contributed equally to this work.}
\author{Alexander Willmes}
\altaffiliation{These authors contributed equally to this work.}
\author{Max Oberl\"ander}
\author{Christian Gorjaew}
\author{Julian D. \surname{Teske}}
\author{Harsh Bhardwaj}
\author{Max Beer}
\author{Eugen Kammerloher}
\affiliation{JARA-FIT Institute for Quantum Information, Forschungszentrum J\"ulich GmbH and RWTH Aachen University, 52074 Aachen, Germany}
\author{Ren\'e Otten}
\affiliation{JARA-FIT Institute for Quantum Information, Forschungszentrum J\"ulich GmbH and RWTH Aachen University, 52074 Aachen, Germany}
\affiliation{ARQUE Systems GmbH, 52074 Aachen, Germany}
\author{Inga Seidler}
\author{Ran Xue}
\affiliation{JARA-FIT Institute for Quantum Information, Forschungszentrum J\"ulich GmbH and RWTH Aachen University, 52074 Aachen, Germany}
\author{Lars R. \surname{Schreiber}}
\altaffiliation{lars.schreiber@physik.rwth-aachen.de}
\author{Hendrik Bluhm}
\altaffiliation{bluhm@physik.rwth-aachen.de}
\affiliation{JARA-FIT Institute for Quantum Information, Forschungszentrum J\"ulich GmbH and RWTH Aachen University, 52074 Aachen, Germany}
\affiliation{ARQUE Systems GmbH, 52074 Aachen, Germany}
\date{\today}

\begin{abstract}
Quantum processor architectures must enable scaling to large qubit numbers while providing two-dimensional qubit connectivity and exquisite operation fidelities. For microwave-controlled semiconductor spin qubits, dense arrays have made considerable progress, but are still limited in size by wiring fan-out and exhibit significant crosstalk between qubits. To overcome these limitations, we introduce the SpinBus architecture, which uses electron shuttling to connect qubits and features low operating frequencies and enhanced qubit coherence. Device simulations for all relevant operations in the Si/SiGe platform validate the feasibility with established semiconductor patterning technology and operation fidelities exceeding $\SI{99.9}{\percent}$. Control using room temperature instruments can plausibly support at least 144 qubits, but much larger numbers are conceivable with cryogenic control circuits. Building on the theoretical feasibility of high-fidelity spin-coherent electron shuttling as key enabling factor, the SpinBus architecture may be the basis for a spin-based quantum processor that meets the scalability requirements for practical quantum computing.
\end{abstract}

\maketitle

The prospect of noisy intermediate scale quantum (NISQ) computing raises high expectations. However, it is likely that a significant part of the foreseen applications will only be accessible via quantum error correction to mitigate errors caused by noise, spurious coupling and crosstalk \cite{Campbell2017}. The resulting overhead leads to a need for millions of physical qubits, which requires highly nontrivial advances compared to today's devices. Electron-spin qubits in semiconductor quantum dots have the unique feature of being directly compatible with industrial CMOS processing \cite{Zwerver2022}. At the level of few-qubit devices, all-electrical operation of single- and two-qubit gates above the error correction threshold have been demonstrated \cite{Yoneda2018, Zajac2018, Watson2018, Xue2019, Petit2020, Xue2022, Noiri2022-1, Takeda2022, Mills2022}. Furthermore, the operation of multi-qubit devices has been shown in several material systems \cite{Lawrie2020, Mortemousque2021, Weinstein2023, Philips2022}. Building on these promising results, the next challenge for semiconductor qubits is scaling-up in two dimensions while simultaneously maintaining high operation fidelities. A key challenge is the short range ($\approx \SI{100}{\nano\meter}$) of the exchange interaction typically used for high-fidelity two-qubit gate operations. Architectures based on direct coupling thus lead to a crowding of gate electrodes and their wiring, referred to as the wiring fan-out problem, as well as significant inter-qubit crosstalk \cite{Undseth2023}.

To address these challenges, dense qubit arrays using crossbar network addressing schemes with reduced wiring density as well as sparse arrays of qubits with integrated classical electronics at cryogenic temperatures have been proposed \cite{Vandersypen2017}. Dense architectures based on crossbar addressing schemes typically apply the same control pulse to many qubits and thus require a challenging level of qubit homogeneity \cite{Li2018}. Tuning the qubit properties with local transistor-based circuits can somewhat ameliorate this issue, but imposes demands on transistor and capacitor size \cite{Veldhorst2017} that are well beyond current capabilities. Sparse arrays of qubits on the other hand require a method for coherent qubit coupling in the order of $\SI{10}{\micro\m}$ \cite{Vandersypen2017, Geck2019, Boter2022}. A possibility is to use electron shuttling, i.e., moving electrons between sites where qubits are manipulated, enabling local exchange-based two-qubit gates. Gate-based electron shuttling has been demonstrated in both GaAs/(Al,Ga)As and Si/SiGe. By implementing Landau-Zehner transitions between adjacent quantum dots in so-called bucket brigade mode, transport of single electrons and coherent transfer of electron spins has already been demonstrated \cite{Baart2016, Flentje2017, Mills2019, Yoneda2021, Noiri2022-2, Zwerver2022-2}. Recently, single electron transport by so-called conveyor-mode shuttling was shown \cite{Seidler2021}, where a quantum dot used to trap the qubit is continuously translated to distant qubit sites, requiring a length-independent number of wires and also less tuning. In a \SI{10}{\micro\meter} long prototype device, charge shuttling in one direction and back across a distance of \SI{19}{\micro\meter} with a fidelity of $\SI{99.7}{\percent}$ has been achieved \cite{Xue2023}.
The concept and feasibility of coherent conveyor-mode electron shuttling was analyzed in detail by Langrock, Krzywda \textit{et al.} \cite{Langrock2023}. The confinement potential is chosen much stronger than the background disorder potential, targeting an adiabatic motion that leaves the electron in the orbital ground state. With a shuttling velocity of a few $\SI{}{\meter\per\second}$, electrons can be transferred fast enough to limit spin dephasing due to $T_2^*$-effects such as charge and hyperfine noise. However, nonadiabatic transitions between different valley and potentially orbital states set an upper bound on the velocity. For a minimal valley splitting of $\SI{20}{\micro\electronvolt}$, a coherent transfer with an error rate below $10^{-3}$ over a distance of $\SI{10}{\micro\meter}$ is predicted for a shuttling velocity of $v=\SI{8}{\meter\per\second}$, which we assume throughout this paper. In a $\SI{1}{\micro\meter}$ Si/SiGe prototype device with natural abundance of Si isotopes (similar to reference \cite{Seidler2021}), spin-coherent shuttling with maximum velocity of $\SI{2.8}{\meter\per\second}$ across an accumulated distance of at least $\SI{2.4}{\micro\meter}$ has been demonstrated. The spin dephasing time of the shuttled electron spin is enhanced by motional narrowing leading to a fidelity of approximately $\SI{99}{\percent}$ for the transfer of a spin quantum state over a nominal shuttling distance of $\SI{560}{\nano\meter}$ \cite{Struck2023}.

\section*{SpinBus architecture and its elements}
\label{sec:Architecture}

\begin{figure*}
    \centering
    \includegraphics[width=\textwidth]{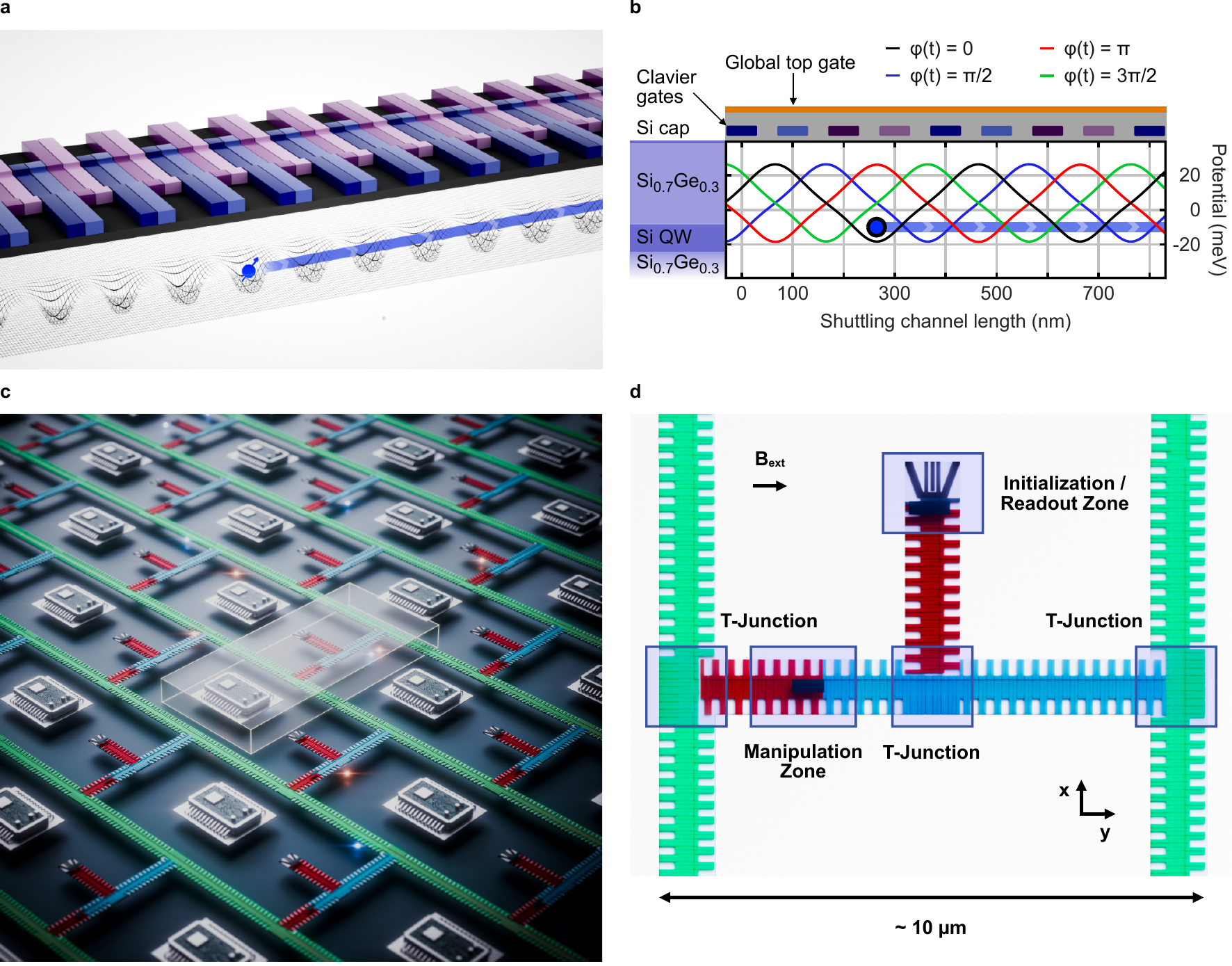}
    \caption{\textbf{Layout and operation of the QuBus device as building block for the SpinBus architecture.} \textbf{a}, 3D visualization of the QuBus device consisting of two lateral screening gates defining a 1D electron channel and periodically connected clavier gates. The four gate sets connected to different control signals $V_{i}$ are color coded. \textbf{b}, Schematic of the Si/SiGe heterostructure providing the quantum well (QW). Linecuts of the traveling potential generated by the gate stack are depicted for four different phases $\varphi \left( t\right)$. The occupied potential minimum is indicated by a red circle. The gate stack is depicted above the potential linecuts. \textbf{c}, The quantum processor chip consists of unit cells tiled like a brick wall. One unit cell is highlighted, the heterostructure is visualized  transparently in most areas and local electronic components are shown symbolically. Unit cells are connected via the green colored shuttling lanes controlled by a signal set shared across unit cells. Red and blue colored shuttling lanes are controlled individually in each unit cell. \textbf{d}, A unit cell consists of three T-junctions for 2D connectivity, an initialization and readout zone, and a manipulation zone. We expect a spatial extent of unit cells in the order of \SI{10}{\micro\meter}.}
    \label{fig:QuBus_Arch}
\end{figure*}

In this manuscript, we present the \textit{SpinBus} architecture, which leverages the conveyor-mode shuttling device named Quantum Bus (QuBus) as used in demonstration experiments \cite{Seidler2021,Xue2023,Struck2023} to connect qubits (Fig. \ref{fig:QuBus_Arch}a). Like established semiconductor qubit devices, the QuBus device employs a stack of electrostatic gates on top of a Si/SiGe heterostructure that confines electrons in the $z$-direction (Fig. \ref{fig:QuBus_Arch}b). Lateral screening gates define a one-dimensional channel in the $xy$-plane, while clavier gates placed above are used to generate moving quantum dots. Every fourth clavier gate is electrically connected, thus eliminating the need for fanning out each individual gate. Four phase-shifted sinusoidal signals $V_{i}$, $i = 1\,...\,4$, applied to the resulting four sets of clavier gates enable a continuous translation of the quantum dots. The signals $V_{i}$ have the form \cite{Seidler2021}
\begin{equation}
    V_{i} = A_{\text{S}} \cos \left( \varphi \left( t \right) - \Delta \varphi_{i} \right) \, .
\end{equation}
Here, $A_{\text{S}}\sim \SI{100}{\milli\volt}$ is the signal amplitude and $\varphi \left( t\right) = 2 \pi f \cdot t$  with frequency $f$ is the phase with phase offset $\Delta\varphi_{i} = \pi/2 \left( i-1\right)$. Hence, the number of required signals is independent of the distance between qubit sites. A DC bias relative to the $V_{i}$ can be applied to Ohmic contacts to adjust the chemical potential.

Based on the QuBus component as coherent link, we propose a layout of tileable unit cells as building blocks for the quantum layer of the SpinBus architecture (Fig. \ref{fig:QuBus_Arch}c). The unit cell (Fig. \ref{fig:QuBus_Arch}d) provides the means for initializing, reading-out and performing gate operations in two specialized zones, i.e., the initialization and readout (IR) and the manipulation zone. Shuttling lanes connect both the operational zones and adjacent unit cells. We anticipate that a length of the shuttling lanes in the order of $\SI{10}{\micro\meter}$ will reflect a reasonable trade-off between shuttling-induced errors and time versus space for wiring and local electronics \cite{Geck2019}. The spatial separation between different manipulation zones and qubits avoids unwanted inter-qubit coupling and helps to address qubits individually, thus avoiding control crosstalk errors. This comes at the cost of shuttling errors, which add to the errors of locally executed gates.

The QuBus geometry is based on the recent demonstration experiments of conveyor-mode shuttling, where a separation of the screening gates by $\SI{200}{\nano\meter}$, a gate width of $\SI{62}{\nano\meter}$ and a gate pitch of $\SI{70}{\nano\meter}$ have been used \cite{Seidler2021}. For the validation of the gate layouts with electrostatic finite-element-method (FEM) models (see Methods \ref{sec:app_es}), we chose a slightly larger gate pitch of $\SI{100}{\nano\meter}$ including a global top gate that can be biased with a separate voltage $V_{\text{tg}} \sim $ \SI{100}{\milli\volt}. For the operation of some elements, micromagnets are placed in suitable locations approximately $\SI{150}{\nano\meter}$ above the quantum well. For magnetostatic modeling (see Methods \ref{sec:app_mag}), we assumed an external in-plane magnetic field $B_{\text{ext}}=\SIrange{20}{50}{\milli\tesla}$ in the $y$-direction.

\begin{figure*}
    \includegraphics[width=\textwidth]{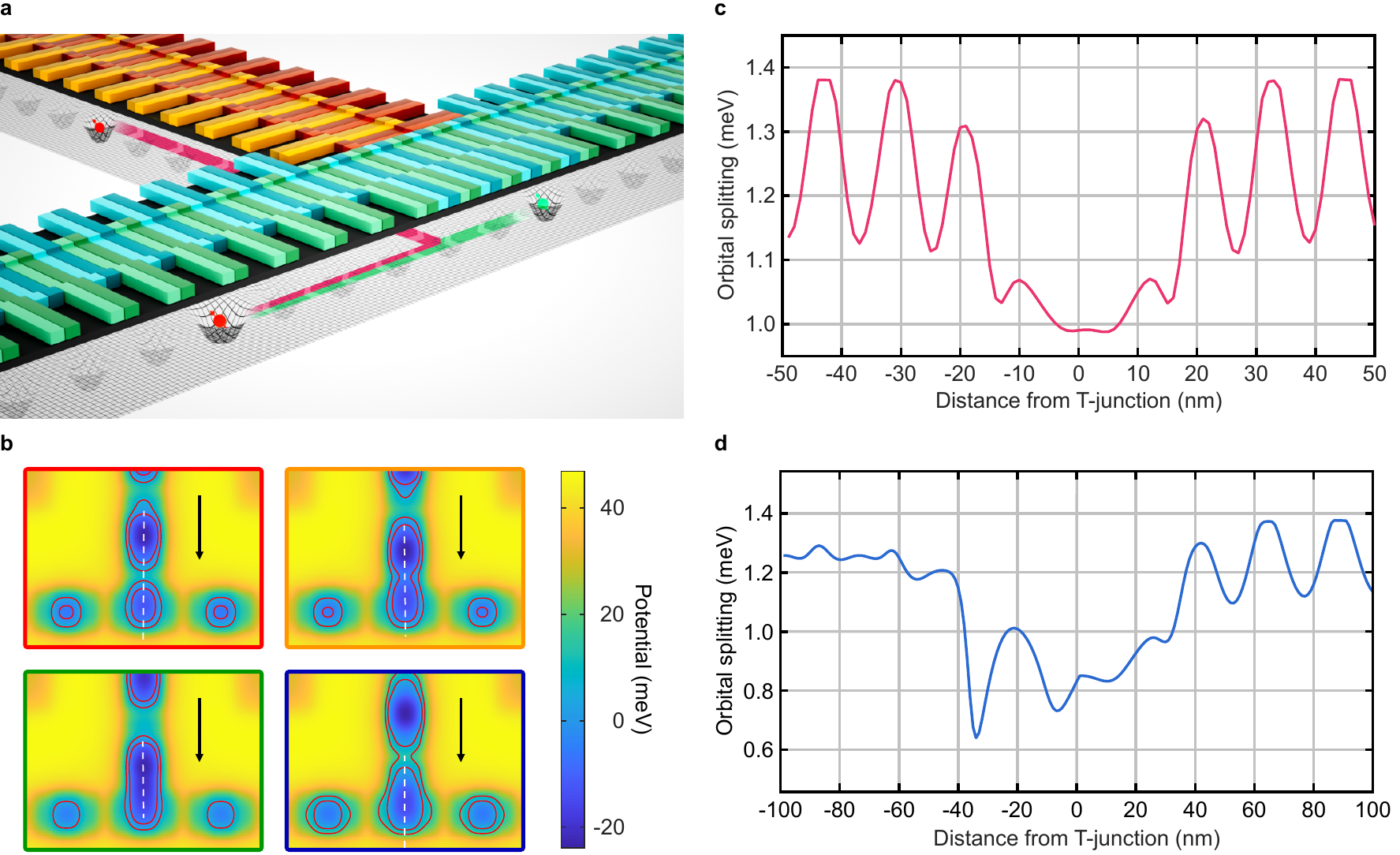}
    \caption{\textbf{Layout and operation of the T-Junction.} \textbf{a}, 3D visualization of the T-junction consisting of two perpendicularly joined QuBus elements. Straight shuttling (red path) and corner shuttling (blue path) are shown. \textbf{b}, 2D potentials for different points in time during corner shuttling. White dashed lines indicate the positions and lengths of the linecuts in Extended Data Fig. \ref{fig-ext:t-junction}a. Arrows indicate the shuttling direction. \textbf{c}, The orbital splitting during straight shuttling is always sufficiently large. \textbf{d}, During corner shuttling, the confinement dips more significantly, but still remains above \SI{0.6}{\milli\eV} near the target range and can be optimized further.}
    \label{fig:t-junction}
\end{figure*}

Two-dimensional connectivity is implemented by a three-way T-junction connecting two perpendicular shuttling lanes (Fig. \ref{fig:t-junction}a). Compared to a four-way junction, gate crowding is reduced and potential shaping simplified. The two supported operations are qubit motion in a straight line (straight shuttling) and around the corner (corner shuttling). Straight shuttling is implemented analogous to normal conveyor-mode operation, with the voltages on the perpendicular branch being constant. For corner shuttling, a quantum dot initially moving along the straight branch is stopped at the intersection and then transferred into the perpendicular branch. Fig. \ref{fig:t-junction}b shows the potentials during the adiabatic transfer using appropriately adjusted voltage pulses. Selected linecuts and the time evolution of the shuttling phases are presented in Extended Data Fig. \ref{fig-ext:t-junction}. For both operations, the transport direction can be inverted by reversing the shuttling pulses. For coherent shuttling, the electron motion should reflect a smooth translation of the potential, rather than tunneling between disorder-induced stationary quantum dots. A useful metric for this requirement is the orbital splitting for the moving quantum dot containing the qubit. If it is similar to the values of \SIrange{1}{2}{\milli\electronvolt} typically found in static quantum dots, one may expect the confinement to be dominated by the shuttling potential and thus a smooth transfer \cite{Langrock2023}. During straight shuttling, the orbital splitting equals or exceeds the threshold at all times (Fig. \ref{fig:t-junction}c). During corner shuttling, the orbital splitting drops to about \SI{0.6}{\milli\electronvolt}, slightly below the conservative target range (Fig. \ref{fig:t-junction}d). We expect that if needed, it can be increased by some combination of optimizing the geometry and pulse shapes, increasing the gate voltages, and dynamically adjusting the screening gate voltage.

\begin{figure*}
    \includegraphics[width=\textwidth]{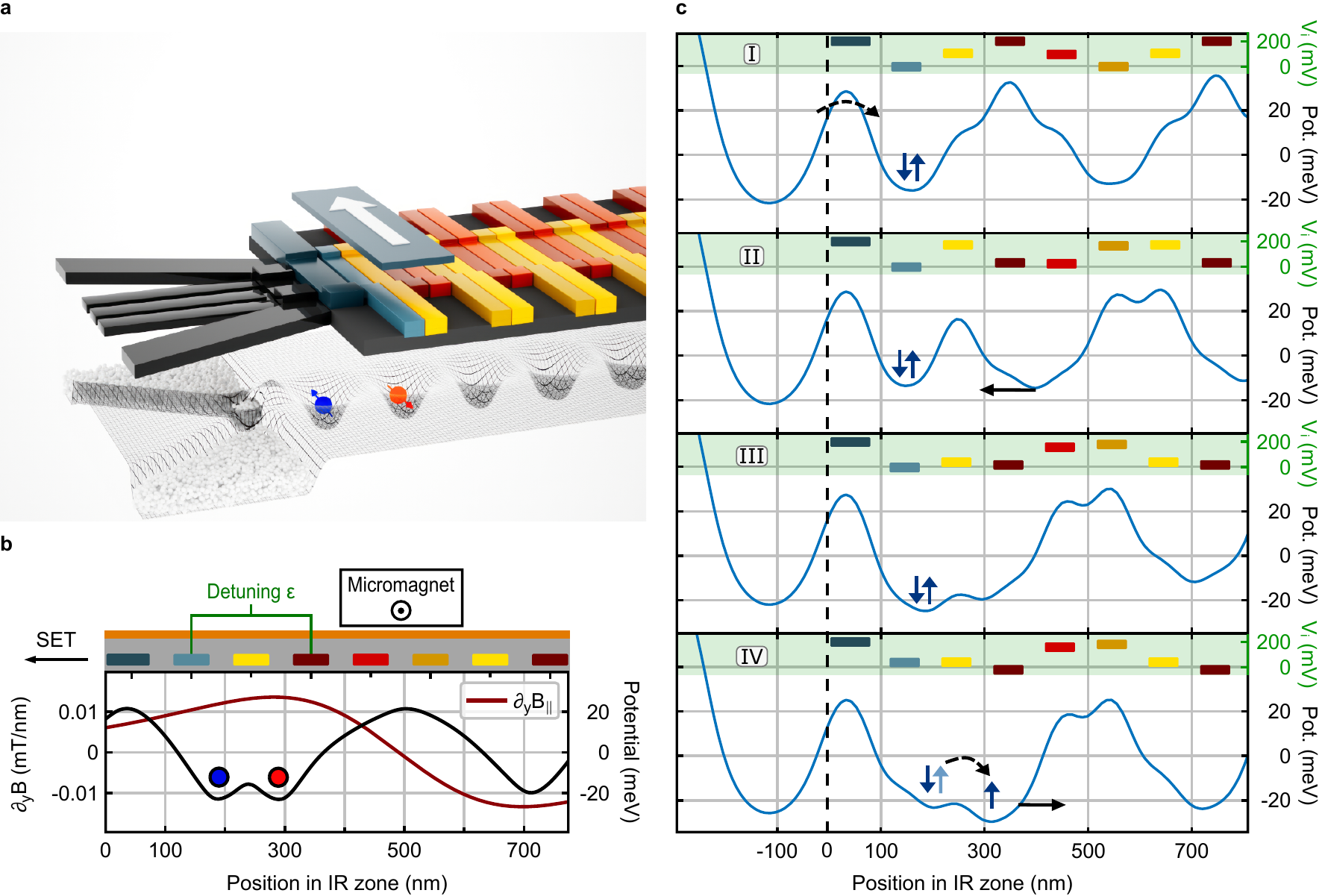}
    \caption{\textbf{Layout and operation of the initialization and readout (IR) zone.} \textbf{a}, 3D visualization of the IR zone consisting of a QuBus element adjacent to an SET, and a micromagnet. Two gates adjacent to four sets of clavier gates are two individually controlled. \textbf{b}, Cross-section including the gate layout showing the schematic double quantum dot potential and simulated magnetic field gradient along the shuttling channel. Red and blue circles represent the positions of two electrons in a double quantum dot configuration. \textbf{c}, Potential linecuts while initializing a qubit using PSB (blue arrows represent the electrons' spin states). The color-coded bars correspond to the gates from panels a and b, and their vertical positions indicate the applied voltage $V_i$. Tunnelling is indicated by dashed black arrows and solid black arrows mark the translation of quantum dots. Step I: loading of a S(2,0) state from the SET into the first quantum dot. Step II: moving a second quantum dot close to the first quantum dot. Step III:  detuned double quantum dot. Step IV: applying a detuning sweep to transfer S(2,0) to $\ket{\uparrow\downarrow}$ followed by a shuttling pulse to inject the qubit into the shuttling channel.}
    \label{fig:IR_zone}
\end{figure*}

The initialization and readout (IR) zone consists of a single electron transistor (SET) tunnel-coupled to a shuttling lane, thus enabling loading and detecting charges (Fig. \ref{fig:IR_zone}a). Ohmic contacts on both sides of the SET provide source and drain reservoirs, and electrons are injected into the shuttling lane via the SET. Besides one plunger and two barrier gates for the SET, we propose two additional individually contacted gates at the beginning of the shuttling lane (Fig. \ref{fig:IR_zone}b). The first controls the tunnel barrier to the SET and the second the potential of the first quantum dot in the QuBus channel. A second moving quantum dot can be controlled independently by the four sets of clavier gates. For qubit initialization and readout, Pauli spin blockade (PSB) in the resulting double quantum dot is utilized to enable simpler and faster readout discrimination than, e.g., spin-selective tunneling \cite{Ono2002, Connors2020}. The required parallel magnetic field gradient $\Delta B_\parallel$ is generated by a micromagnet placed directly above the shuttling lane adjacent to the SET. The initialization sequence follows standard procedures and is presented in Fig. \ref{fig:IR_zone}c. It starts with loading two electrons into a first quantum dot (step I), forming a tunnel-coupled double quantum dot configuration (step II) while the second quantum dot is kept at a sufficiently higher potential during the adjustment of the inter-dot tunnel barrier (step III) to remain in a S(2,0) state. Sweeping the detuning $\epsilon$ transfers the S(2,0) state to a (1,1) configuration, where the gradient magnetic field splits the T$_0$ and S(1,1) into $\ket{\uparrow\downarrow}$ and $\ket{\downarrow\uparrow}$ (step IV). Thus, the $\ket{\downarrow\uparrow}$ will be occupied if the detuning is pulsed adiabatically with respect to orbital, spin and valley excitations, but including a short diabatic sweep over the ST$_-$-crossing. Lastly, the spin-up state is shuttled away to be used as qubit. The spin-down electron can be kept in the first quantum dot as reference spin state for later readout. The corresponding time traces for the shuttling phase $\varphi \left( t\right)$ and detuning are shown in Extended Data Fig. \ref{fig-ext:IR_zone}. To implement readout, the initialization sequence is reversed and PSB is employed to determine the qubit's state. Any established method for SET readout can be used, though we speculate that baseband readout with cryogenic transistors \cite{Vink2007, Tracy2016, Curry2019} will yield the best performance-complexity trade-off.
 
\begin{figure*}
    \includegraphics[width=\textwidth]{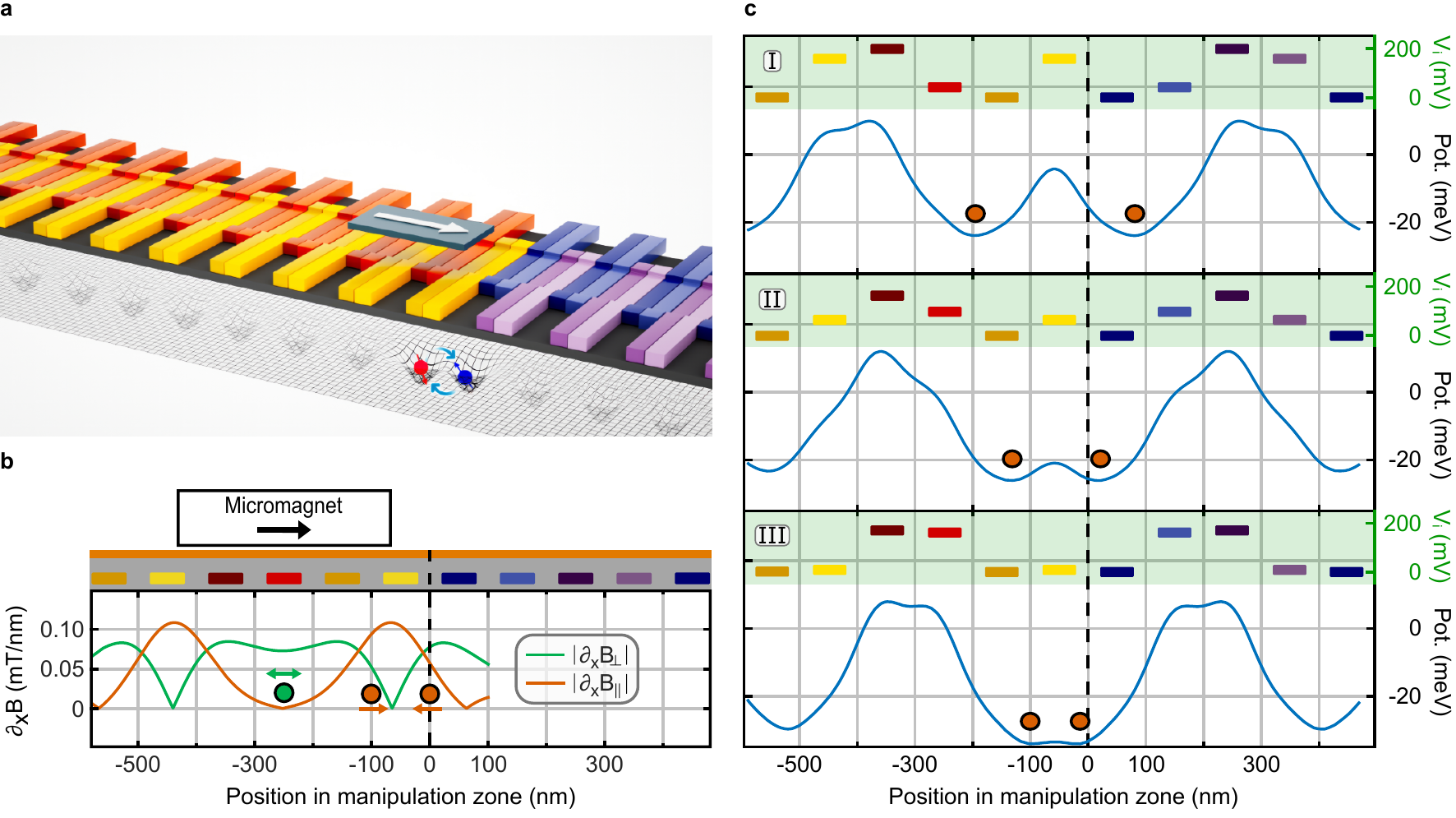}
    \caption{\textbf{Layout and operation of the manipulation zone.} \textbf{a}, 3D visualization of the manipulation zone  consisting of two joined QuBus elements and two micromagnets. Shown is an exemplary two-qubit operation. \textbf{b}, Cross-section including the gate layout showing the required magnetic field gradients for single- and two-qubit gates along the manipulation zone. The green circle shows the position of the qubit during the single-qubit gate operation and is driven periodically in the region of a large perpendicular magnetic field gradient $|\partial_x B_\perp|$. The two orange circles indicate the positions of the qubits during the two-qubit gate operations. Both are pushed together at the location of a large parallel magnetic field gradient $|\partial_x B_\parallel|$. \textbf{c},  Potential linecuts showing the smooth formation of a tunnel-coupled double quantum dot potential appropriate for two-qubit operations as the two translated quantum dots approach the center of the manipulation zone. The color-coded bars correspond to the gates from panels a and b, and their vertical positions indicate the applied voltage $V_i$.}
    \label{fig:Manipulation_zone}
\end{figure*}

Single- and two-qubit gate operations are performed in the manipulation zone, which is formed by joining two shuttling lanes (Fig. \ref{fig:Manipulation_zone}a). Two independent QuBus elements enable sufficient control over both detuning and tunnel coupling of a double quantum dot potential formed at the junction. Two micromagnets provide the necessary magnetic field gradients (Fig. \ref{fig:Manipulation_zone}b). For two-qubit gates and single-qubit gates, a micromagnet is placed off-center from the junction above one QuBus element. An additional micromagnet for single-qubit gates on the other side of the junction is located above the other QuBus element at a sufficient distance to avoid compromising the longitudinal field gradient at the junction. Thus, the manipulation zone allows performing single-qubit gates on two qubits independently. Shuttling the qubits $\SI{200}{\nano\meter}$ to $\SI{500}{\nano\meter}$ away from the junction for single-qubit gates can avoid crosstalk. Single-qubit gates are implemented by electric-dipole spin resonance (EDSR), in which an effective oscillatory transverse magnetic field for driving Rabi oscillations is generated by displacing the electron in a perpendicular magnetic field gradient. Unlike conventional EDSR manipulation, where the electron position oscillates typically up to one nanometer \cite{Yoneda2018}, we propose a shuttling-mode EDSR building on the capability of moving the electron over arbitrary distances. For high fidelities, we estimate an oscillation amplitude in the order of $\SI{10}{\nano\meter}$ to be a good choice. The larger amplitude allows the use of significantly weaker magnetic field gradients, which reduces the sensitivity to charge noise. 

For electron spin qubit platforms utilizing micromagnets, the natural choice for the implementation of CNOT-like two-qubit gates (see Methods \ref{sec:app_CNOT}) is the controlled-phase (CPHASE) gate based on the exchange interaction $J \left(t\right)$ between two tunnel-coupled quantum dots \cite{Loss1998, Meunier2011, Russ2018}, which is switched adiabatically with respect to a Zeeman energy difference $\Delta E_{\text{Z}}$ between the two quantum dots. This configuration is achieved by shuttling both electrons to the junction at the center of the manipulation zone (Fig. \ref{fig:Manipulation_zone}c) with pulses as shown in Extended Data Fig. \ref{fig-ext:Manipulation_zone} while maintaining zero detuning. The control of the exchange coupling via the inter-qubit distance while maintaining zero detuning essentially amounts to barrier control, which features a lower charge noise sensitivity compared to controlling the exchange interaction via the detuning \cite{Watson2018, Petit2022}. Fig. \ref{fig:Manipulation_zone}c shows the simulated potentials during the formation of a double quantum dot. The separation and barrier height during the two qubit gates are similar as in conventional quantum dot structures, thus validating the robustness of the procedure with respect to disorder. The absence of tunnel coupling to other sites further increases this robustness in comparison to arrays with multiple quantum dots.

\section*{Fidelity of quantum operations}
\label{sec:Fidelities}
To estimate the achievable performance, we simulated the dynamics of each quantum operation using the simulation package qopt\cite{Teske2022}, including an optimization of the control pulses (see Methods \ref{sec:app_ham}). The fidelities were computed based on a noise model including quasistatic nuclear spin noise affecting the Zeeman splitting as well as quasistatic and white charge noise with amplitudes extracted from past experiments \cite{Dial2013, Kranz2020, Yoneda2018}. With appropriate calibration, the combination of quasistatic and white charge noise can serve as a conservative proxy for $1/f$-noise typically found in real devices. We included coupling of the charge noise to the qubit via the detuning affecting the exchange coupling as well as via position fluctuations. The latter affect the single spin dynamics due to the magnetic field gradient as well as the exchange coupling at zero detuning. This noise model covers the effects we consider as experimentally most relevant and was shown to be in good agreement with experimental results \cite{Cerfontaine2020}. For the initialization and readout procedure, we identified fast charge noise as the main limiting factor and obtained fidelities above $\SI{99.9}{\percent}$ if parasitic inter-dot orthogonal magnetic field gradients remain sufficiently small (see Methods \ref{sec:app_mag}). To evaluate single-qubit operations, we applied a sinusoidal shuttling EDSR-pulse in resonance with the Zeeman splitting to a qubit model with spin and valley degree of freedom. We identified fast charge noise causing position fluctuations as the dominating noise contribution and find fidelities exceeding $\SI{99.9}{\percent}$ as long as the valley splitting is greater than $\SI{30}{\micro\electronvolt}$ and does not exhibit an exceptionally strong variation. For two-qubit gates, the relevant infidelity contribution arises from quasistatic position noise affecting the exchange interaction and we obtain a fidelity of $\SI{99.9}{\percent}$.

\section*{Operating concept and system complexity}
\begin{figure*}
  \includegraphics[width=\textwidth]{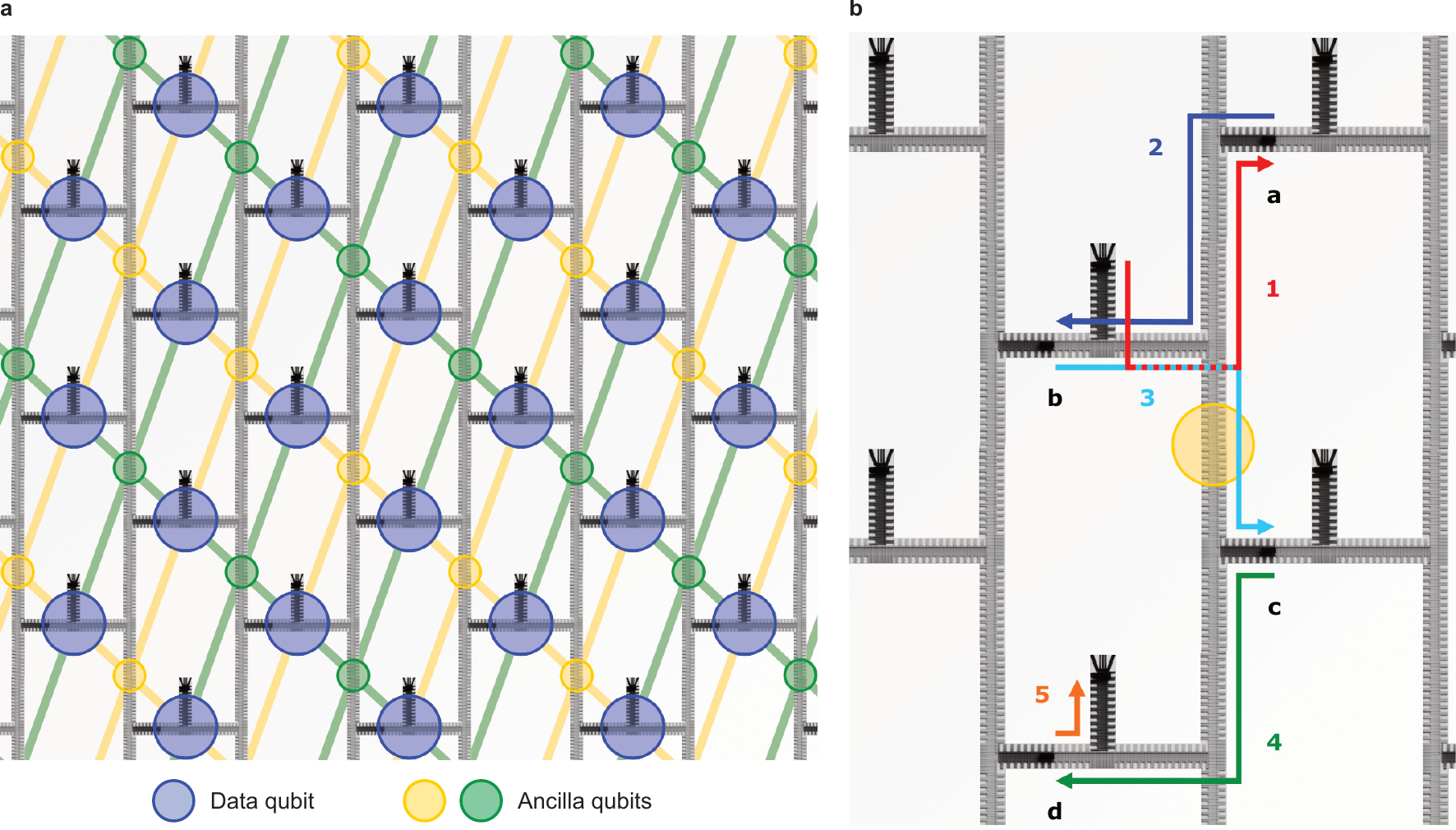}
  \caption{\textbf{Exemplary surface code implementation in the SpinBus architecture.} \textbf{a}, Mapping of data and ancilla qubits to unit cells. Each diagonal line segment represents a required qubit interaction. \textbf{b}, Shuttling paths for an ancilla qubit in order to implement an $\hat{X}$ stabilizer. Associated data qubits are designated with \textit{a}, \textit{b}, \textit{c}, \textit{d}. Arrows mark the shuttling paths to all involved manipulation zones, whereas the numbers indicate the order of operations.}
  \label{fig:AlgImplementation}
\end{figure*}

The two-dimensional array of the architecture is well suited for the implementation of surface codes, which can be considered as the mainstream concept for quantum error correction \cite{Campbell2017}, as well as NISQ algorithms. As exemplary operation, we show the elementary surface code gate sequence in Fig. \ref{fig:AlgImplementation}, requiring a square array of qubits with nearest-neighbor coupling. Every second qubit serves as data qubit storing quantum information, and every other one as ancilla qubit, each detecting one of two possible types of errors called $\hat{X}$ and $\hat{Z}$ stabilizers \cite{Fowler2012}. As each manipulation zone can simultaneously operate two qubits, each unit cell is identified with one data qubit highlighted in blue and one adjacent ancilla qubit highlighted in green and yellow, respectively (Fig. \ref{fig:AlgImplementation}a). An error detection cycle consists of initializing the ancilla qubits, CNOT gates with the four adjacent data qubits, which we choose as stationary, and subsequent readout of the ancilla qubits. Realizing such a cycle in the SpinBus architecture requires the shuttling of ancilla qubits to and from different manipulation zones between local gate operations (Fig. \ref{fig:AlgImplementation}b).

As the wiring density can be a main limiting factor for the size of processor achievable with a given integration approach, we present an estimate of the number of required signals based on an economical operating strategy detailed in Supplementary information \ref{sec:sup_wiring}. Considering shuttling signals (also used for qubit control) and additional local AC and DC signals of the IR zone and the screening gates, a quantum processor chip with $N$ unit cells requires $14\,N+4$ AC and $3\,N+4$ DC signals. While there are no inherent scaling limitations to our architecture at the quantum layer, the wiring requirements have to be compatible with cryostat wiring, packaging and back-end-of-line (BEOL) technology. We estimate that currently available wiring solutions in cryostats of about 1,000 coaxial cables\cite{Bluefors} are the most limiting factor and can accommodate a quantum processor chip with $9 \times 8 = 72$ unit cells. This corresponds to 144 simultaneously operable qubits if two qubits per unit cell are loaded. Storing additional qubits away from manipulation zones can further increase the qubit number.

\section*{Conclusion}
\label{sec:Conclusion}
In summary, we have detailed a concept to leverage electron shuttling for the realization of a semiconductor-based quantum processor with 2D coupling, as required for quantum error correction based on the surface code. To validate the feasibility, we performed electrostatic simulations for all device layouts and modes of operation. For the estimation of operation fidelities, we used realistic noise models and obtained fidelities for single- and two-qubit gates exceeding $\SI{99.9}{\percent}$. The fabrication is possible with present-day industrial semiconductor processing. Furthermore, the architecture is compatible with established packaging and wiring techniques such as back-end-of-line (BEOL) via fabrication and flip-chip bonding. While we considered an implementation in Si/SiGe, the SpinBus architecture can potentially be transferred to other types of gate-defined semiconductor qubits.

Our architecture proposal features a number of strengths, but it clearly hinges on the theoretically predicted feasibility of spin-coherent electron shuttling. While first experiments on spin-coherent transport are promising, an implementation with high fidelity and mitigating low values of the valley splitting in Si/SiGe (see Supplementary information \ref{sec:sup_valley} for details) will be an essential next step. Reaching the projected fidelities and required yield could quickly put semiconductor qubits on the map for NISQ-type quantum computing. The combination with cryoelectronic control systems, which is facilitated by the variable qubit spacing, robust coherence of semiconductor qubits, the purely capacitive load of gate electrodes and the relatively low operating frequency could carry to much larger systems, eventually enabling error corrected quantum computing. As outlined in Supplementary information \ref{sec:sup_scale}, recent advances in cryoelectronics and packaging provide concrete perspectives on how this goal can be tackled.

\appendix

\makeatletter
\renewcommand \thesubsection{M\@arabic\c@subsection}
\makeatother

\section*{Methods}
\subsection{Electrostatic simulations and orbital splitting}
\label{sec:app_es}
For the calculation of the electrostatic potentials, we employed finite-element-method (FEM) simulations using COMSOL Multiphysics$^{\text{\textregistered}}$. For each operational element, as shown in Figs. \ref{fig:t-junction}a, \ref{fig:IR_zone}a, \ref{fig:Manipulation_zone}a of the main text, we solved Poisson's equation:
\begin{equation}
    - \nabla (\epsilon(r)\nabla\Phi)=\frac{1}{\epsilon_0} \rho \, ,
\end{equation}
with the electrostatic potential $\Phi$, charge density $\rho$, dielectric constant of the sample $\epsilon(r)$ and the vacuum permittivity $\epsilon_0$. Dirichlet boundary conditions corresponding to the applied voltages were imposed at metallic gates. As the structure is intended to be filled with dilute electrons representing qubits whose behavior will be fully governed by the electrostatic potential in their absence, their charge was not included in $\rho$. We used the linearity of the model to simplify the variation of the applied voltages $V_i$ by calculating basis potentials $\Phi$ of each gate $i$ separately and combining the resulting total potential
\begin{equation}
    \Phi = \sum_i V_i \Phi_i \, .
\end{equation}
Specifically, $\Phi_i$ is the potential for gate $i$ set to $\SI{1}{\volt}$ with all others at $\SI{0}{\volt}$. This superposition approach is justified in regions where no or only very few electrons are present. In the IR zone, however, one needs to take the reservoir's contribution to $\rho$ into account. To do so, we first used the Thomas-Fermi approximation and solved the Poisson equation self-consistently, assuming a depleted 2DEG in the channel and SET, for a specific gate voltage configuration that leads to the intended occupation of the reservoirs. To simplify fine tuning of the gate voltages via the superposition approach, we subsequently modeled the reservoirs analogous to metallic gates, thus assuming perfect screening. The position of these gates was obtained from the region of nonzero charge density of the initial Thomas-Fermi solution. This neglects a change of the reservoir region in response to gate voltages and introduces a small error as the gradual screening by the 2DEG in the reservoirs is replaced by a hard boundary conditions. As the reservoirs are relatively far from the region of interest, these approximations are compatible with our goal of demonstrating the feasibility o create an appropriate potential.
The potential energies shown in figures are referenced to the conduction band edge and given as $V=-e \Phi$.

For quantifying the effect of the variations in confinement in the T-junction (Fig. \ref{fig:t-junction}), we calculated the orbital splitting for the simulated potential of the quantum dot confining the qubit by solving the time-independent Schr\"odinger equation in two dimensions for each time step.

\subsection{Micromagnet design}
\label{sec:app_mag}
Considering the requirements for the gate operations, we identified suitable dimensions for Cobalt micromagnets which provide the necessary field gradients. The resulting geometries and corresponding gradients are summarized in Table \ref{tab:microm_params_150nm}. Using thin layers ensures sufficient remanent magnetization when operating at low external magnetic fields, which we substantiated with OOMMF\cite{Donahue1999}-simulations using material parameters from \cite{Neumann2015}. Note that the perpendicular field gradient $\Delta B_\perp$ for two-qubit gates arising from the magnet geometry is neither required nor harmful. As it is weaker than in the single-qubit zone and the gate duration is comparable, resulting relaxation errors are expected to be negligible.

\begin{table}[ht]
    \centering
    \begin{tabular}{c|c|c|c}
    Case & dimension / $\SI{}{\nano\meter^3}$ & $\Delta B_\perp$ $\vert$ $\partial B_\perp$ & $\Delta B_\parallel$ $\vert$ $\partial B_\parallel$\\
    \hhline{=|=|=|=}
    \multirow{1}{*}{IR}  & $\SI{700}{}\times\SI{200}{}\times\SI{20}{}$ & $<\SI{0.3}{\milli\tesla}$ & \SI{1.3}{\milli\tesla} \\ \cline{1-4}
    \multirow{1}{*}{SQG}  & \multirow{2}{*}{$\SI{400}{}\times\SI{200}{}\times\SI{20}{}$} & \SI{0.075}{\milli\tesla\per\nano\meter} & $<\SI{0.01}{\milli\tesla\per\nano\meter}$\\ \cline{1-1}\cline{3-4}
	\multirow{1}{*}{TQG}  &  & $\SI{4}{\milli\tesla}$ & $\SI{8.7}{\milli\tesla}$\\
    \end{tabular}
    \caption{\textbf{Dimensions of the micromagnets for the IR zone and manipulation zone, respectively, and associated magnetic field gradients.} IR: initialization and readout. SQT: single-qubit gates. TQG: two-qubit gates.}
    \label{tab:microm_params_150nm}
\end{table}

\subsection{Operation fidelities}
\label{sec:app_ham}
To verify the feasibility of the architecture, we performed quantum dynamic simulations of each quantum operation using the simulation package qopt\cite{Teske2022}. We calculated the quantum dynamics by solving the time-dependent Schr\"odinger equation for adequate model Hamiltonians to identify simple control pulses for initialization and readout, single-qubit gates and two-qubit gates. To extract meaningful fidelities, we included realistic noise values from past experiments \cite{Dial2013,Kranz2020,Yoneda2018}. All simulations were performed assuming the g-factor of Si.

For the initialization and readout procedure, we simulated a linear ramping pulse which converts between the $S(2,0)$ and $\ket{\downarrow\uparrow}$-state by sweeping the potential detuning $\epsilon$ of a double quantum dot adiabatically, besides a jump over the avoided $ST_-$-crossing. We utilized a Hamiltonian truncated to the relevant three-state basis of $\{\ket{T_0},\ket{S},\ket{T_-}\}$,
\begin{align}
	H = \begin{pmatrix}
		0 		 		& \Delta B_\parallel / 2 			& 0 \\
		\Delta B_\parallel / 2 	& -J(\varepsilon) 			& \Delta B_\perp / (2 \sqrt{2}) \\
		0 				& \Delta B_\perp / (2 \sqrt{2}) & -B_\parallel
	\end{pmatrix}\,, \label{eq:hamiltonian_init_readout}
\end{align}
taking the Zeeman splitting ($B_\parallel$), parallel ($\Delta B_\parallel$) and orthogonal ($\Delta B_\perp$) field differences between two dots from micromagnet simulations and experimental data for the exchange energy $J(\varepsilon)$ from Dial \textit{et al.} \cite{Dial2013}. Including further fast charge noise of $\sqrt{S_{\varepsilon}}=\SI{0.02}{\nano\electronvolt\per\sqrt{\hertz}}$ (adopted from Dial \textit{et al.} \cite{Dial2013} assuming a lever arm of 0.1) and optimizing a jump in $\epsilon$ at the avoided crossing of $\ket{T_-}$ and $\ket{S}$ induced by unintentional orthogonal field gradients gives target state fidelities exceeding $\SI{99.9}{\percent}$ when choosing pulse lengths $t\sim\SI{200}{\nano\second}$, fields of $B_\parallel\geq \SI{20}{\milli\tesla}$, $\Delta B_\parallel\sim \SI{1}{\milli\tesla}$ and a parasitic a inter-dot orthogonal magnetic field gradient $\Delta B_\perp \lessapprox \SI{0.3}{\milli\tesla}$. The separation of the electrons, which is well established, was assumed to occur perfectly adiabatically without thermal or dynamic excitation.\par 
For single-qubit EDSR, spin and valley degree of freedom were considered in the Hamiltonian
\begin{align}
H&=\frac{1}{2}B_\parallel(x)\sigma_z+\frac{1}{2}B_\perp(x)\sigma_x+\Delta_{\mathrm{VS,x}}(x)\tau_x+\Delta_{\mathrm{VS,y}}(x)\tau_y\nonumber\\
&+\kappa_{\mathrm{SVC,x}}\tau_x\otimes\sigma_z+\kappa_{\mathrm{SVC,y}}\tau_y\otimes\sigma_z\label{eq:hamiltonian_single_qubit_gate}
\end{align}
with $\sigma_i$ and $\tau_i$ denoting Pauli matrices on spin and valley space, respectively, $\Delta_{\mathrm{VS}}=\Delta_{\mathrm{VS,x}}+i\Delta_{\mathrm{VS,y}}$ is a complex matrix element describing the coupling of the two lowest near-degenerate valley states in silicon \cite{Friesen2007} as a function of the electron position $x$ and $\kappa_{\mathrm{SVC,i}}=\SI{0.01}{\micro\electronvolt}$ parametrizes a $g$-factor variation between the valley states. EDSR-pulses as enveloped sinusoidal drives were then optimized for resonance frequency with the software framework \texttt{qopt} \cite{Teske2022} and evaluated with respect to a process fidelity in the sense of Wood \textit{et al.}\cite{Wood2018}, taking into account leakage as valley excitations entail phase errors in subsequent operations \cite{Langrock2023}. From the simulations we identified fast charge noise as the dominating noise contribution, which we modeled as an effective positional fluctuation of $\sqrt{S}=\SI{0.1}{\femto\meter\per\sqrt{\hertz}}$ based on the experiments of Dial \textit{et al.} and Yoneda \textit{et al.} \cite{Yoneda2018,Dial2013}. Taking into account the design choice of weaker magnetic gradients $\partial B_\perp\sim\SI{.1}{\milli \tesla \per \nano \meter}$, our results indicate displacement amplitudes of around $\SI{20}{\nano\meter}$ (peak-to-peak) as a viable operation point to uphold a Rabi frequency near $\SI{10}{\mega \hertz}$. This displacement amplitude constitutes a trade-off between achievable Rabi frequency and decoherence due to leakage on the valley space from potentially non-uniform valley splitting over the increased traveling distance of the electron during the pulse compared to previous experiments. The possibility to reach a fidelity of $\SI{99.9}{\percent}$ was found to be strongly correlated with the presence of a valley splitting $\gtrapprox\SI{30}{\micro\electronvolt}$ with sufficient spatially uniformity. Employing a model for $\Delta_{\mathrm{VS}}$ incorporating alloy disorder effects on the valley splitting recently proposed in Wuetz \textit{et al.}\cite{Wuetz2022} then yields $>80\%$ probability for fidelities $>\SI{99.9}{\percent}$ in the initial environment of the electron inside the manipulation zone under the assumption of $\expval{E_{\mathrm{VS}}}=2\expval{\abs{\Delta_{\mathrm{VS}}}}=\SI{100}{\micro\electronvolt}$, which is a conservative value compared to the current state of the art \cite{Hollmann2020}.  Adjustment of the electron position within the range of manipulation zone, and therefore its valley environment, can further be utilized to circumvent spots with pathological behavior of the valley splitting that compromises the performance.\par
Two-qubit interaction was examined by doubling the Hamiltonian Eq. (\ref{eq:hamiltonian_single_qubit_gate}), additionally introducing a dot distance $d$ dependent exchange energy $J(d)$, and optimizing a CZ-class gate from local invariants \cite{Watts2015} with an adiabatic pulse bringing together both electrons. The exchange energy $J(d)$ was calculated by solving the two-electron Schr\"odinger equation in one spatial dimension along the channel for each potential configuration of the shuttle pulse. $d$ was obtained as the separation between the minima of the double well potential in the two QuBus elements adjacent to the manipulation zone. Physically, both gate voltage fluctuations as well as charge noise contribute to fluctuations in $J$. Rather than modelling these independently, which is difficult to calibrate based on experiments anyway, we introduced an effective noise in $d$. The relevant infidelity contribution then arose from quasistatic modifications affecting the exchange interaction, which we estimated conservatively from  Yoneda \textit{et al.} \cite{Yoneda2018}  to $\sigma_d=\SI{10}{\pico\meter}$. This mathematical parametrization of $J$-noise in terms position fluctuations can be expected to give a reasonable estimate because position variations directly translate to a change in the barrier height and width, which is the main factor for $J$ in the assumed barrier control mode. The assumed magnetic field gradient of $\partial B_\parallel\sim\SI{.1}{\milli \tesla \per \nano \meter}$ was found suitable to realize entangling dynamics of CZ-class interaction on timescales of $t_{\mathrm{int}}\sim\SI{50}{\nano\second}$ with a fidelity exceeding $\SI{99.9}{\percent}$ conditional on coherent electron shuttling capabilities requiring sufficient ($\gtrapprox\SI{30}{\micro\electronvolt}$) valley splitting.\par

\subsection{CNOT gate synthesis}
\label{sec:app_CNOT}
\begin{figure}
    \includegraphics[width=0.5\textwidth]{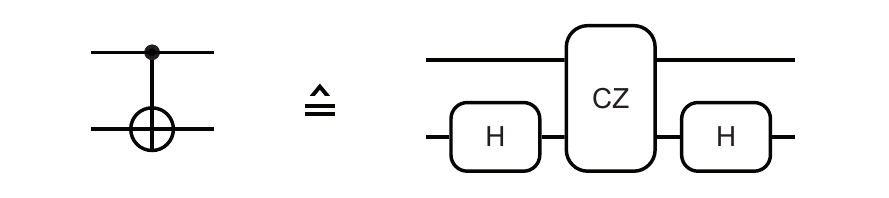}
    \caption{\textbf{CNOT gate synthesis} A CNOT gate is synthesized by a CZ gate, including Hadamard-like operations to account for necessary single-qubit operations (e.g., $\hat{z}$ rotations).}
    \label{fig-meth:CNOT}
\end{figure}

CNOT gates are synthesized from CZ gates\cite{Mills2022} (Fig. \ref{fig-meth:CNOT}), since for electron spin qubit platforms utilizing micromagnets, the natural choice for the implementation of CNOT-like two-qubit gates is the controlled-phase (CPHASE) gate. It requires a Zeeman energy difference $\Delta E_{\text{Z}}$ and an adiabatically switched exchange interaction $J \left(t\right)$ between two tunnel-coupled quantum dots \cite{Loss1998, Meunier2011, Russ2018}. The actual gate operation is based on adiabatically turning on the exchange interaction $J \left(t\right)$, which shifts the energy levels of the antiparallel spin states in such a way that they acquire additional phases. Applying an exchange pulse for a duration $\tau=\pi\hbar/J$ combined with appropriately calibrated single-qubit gates \cite{Vandersypen2001, Watson2018, Petit2020, Mills2022} allows the implementation of a controlled-Z (CZ) gate or a CNOT gate to realize a universal gate set \cite{Mills2022}. 

\section*{Acknowledgements}
This work has been funded by the European Research Council (ERC) under the European Union’s Horizon 2020 research and innovation program (Grant agreement No. 679342), by the German Research Foundation (DFG) under Germany’s Excellence Strategy - Cluster of Excellence Matter and Light for Quantum Computing" (ML4Q) EXC 2004/1-390534769, by the Federal Ministry of Education and Research (Germany), Funding reference number: 13N15652 and  via Project Si-QuBus within the QuantERA ERA-NET Cofund in Quantum Technologies implemented within the European Union’s Horizon 2020 research and innovation program.

\section*{Author contributions}
M.K., A.W., I.S., L.R.S. and H.B. conceived the device layouts. M.K., A.W., H.Bh., M.B., E.K., I.S. and R.X. implemented and carried out electrostatic simulations. M.O., C.G. and J.D.T. carrier out the dynamics simulations for each quantum operation. M.K., A.W. and H.B. conceived the operating strategy. M.K., A.W., R.O. and H.B. developed the wiring concept. H.B. and L.R.S. provided guidance to all authors. M.K., A.W., M.O., J.D.T. and H.B. wrote the manuscript.

\section*{Competing interests}
M.K., A.W., M.O., J.D.T., H.Bh., E.K., I.S., R.X., L.R.S. and H.B. are co-inventors of patent applications that cover conveyor-mode shuttling and various aspects of the SpinBus architecture. L.R.S. and H.B. are founders and shareholders of ARQUE Systems GmbH. The other authors declare no competing interest.

\clearpage

\section*{Extended data}
\makeatletter
\renewcommand{\figurename}{\textbf{Extended Data Fig.}}
\setcounter{figure}{0}
\makeatother

\begin{figure*}
    \includegraphics[width=\textwidth]{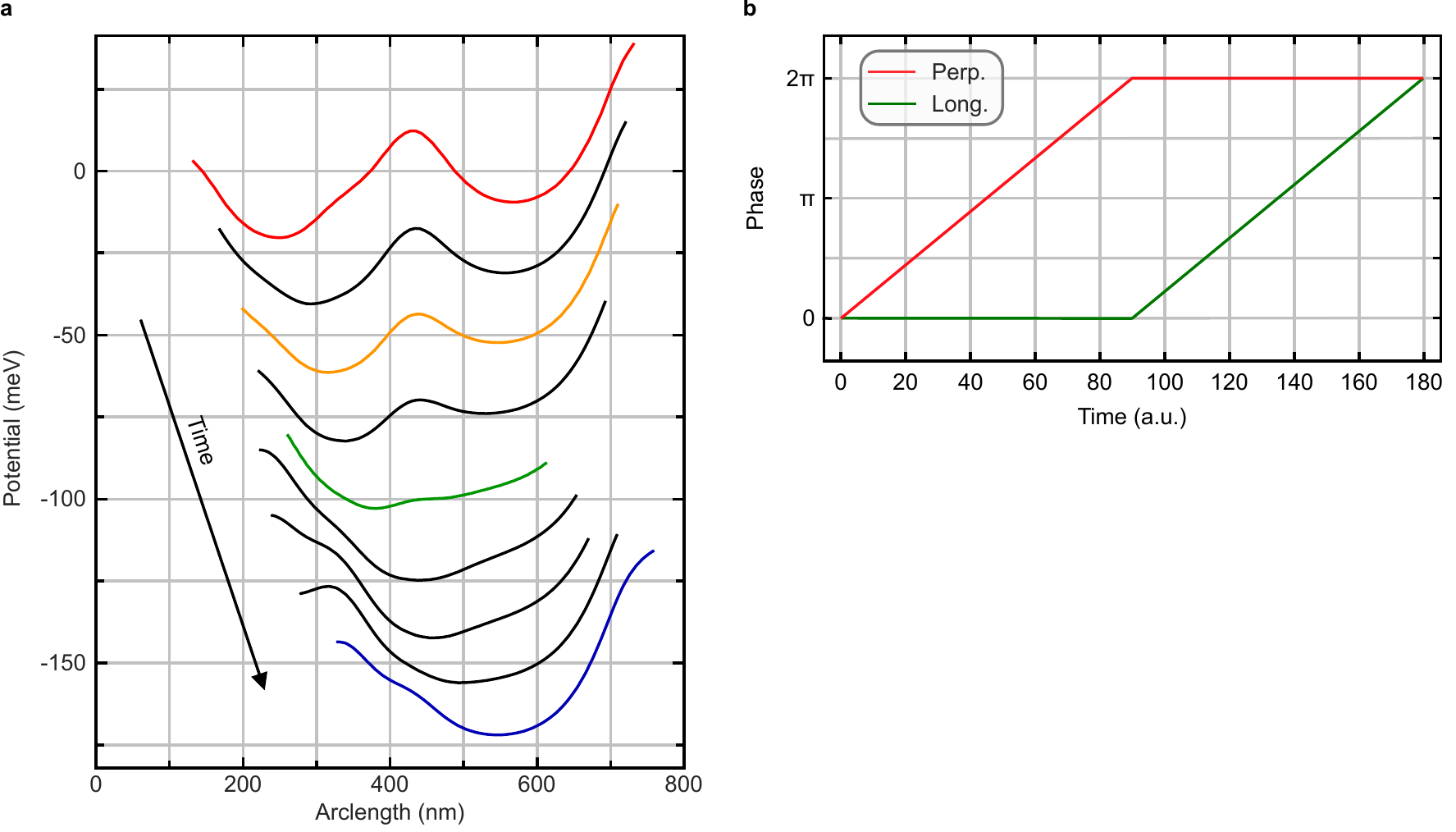}
    \caption{\textbf{Potential linecuts and shuttling phases during T-junction operation}. \textbf{a}, Linecuts of the potential for different times during corner shuttling. Colors correspond to the frames in panel Fig. \ref{fig:t-junction}b. Successive linecuts are shifted for clarity. \textbf{b}, Time evolution of shuttling phases during corner shuttling for the initial perpendicular (red) and target longitudinal (green) shuttling element.}
    \label{fig-ext:t-junction}
\end{figure*}

\begin{figure*}
    \includegraphics[width=0.5\textwidth]{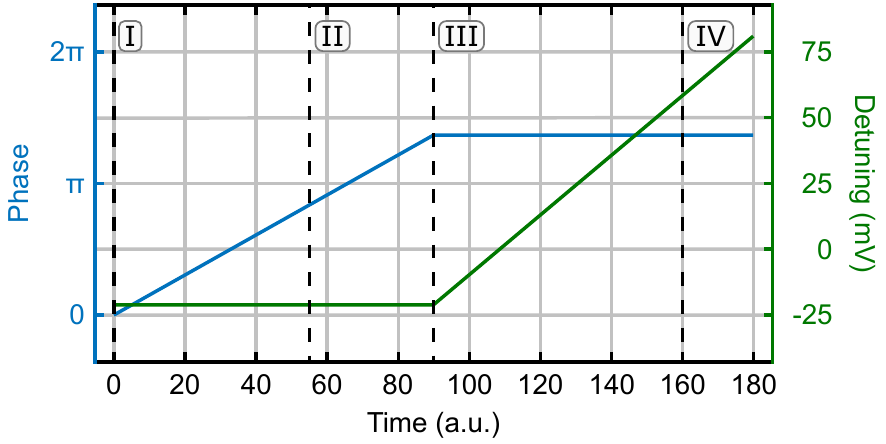}
    \caption{\textbf{Time evolution of the shuttling phase $\varphi(t)$ and interdot-detuning during initialization.} Dashed lines indicate the points in time of the potential linecuts in Fig. \ref{fig:IR_zone}c.}
    \label{fig-ext:IR_zone}
\end{figure*}

\begin{figure*}
    \includegraphics[width=0.5\textwidth]{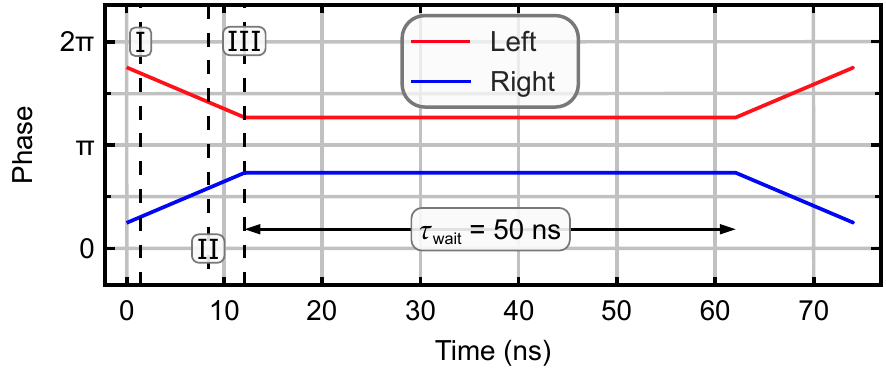}
    \caption{\textbf{Time evolution of shuttling phases during a two-qubit operation.} Included is a wait time of $\tau_{\text{wait}}=\SI{50}{\nano\second}$ for the actual two-qubit gate. Dashed lines indicate the points in time of the potential linecuts in Fig. \ref{fig:Manipulation_zone}c.}
    \label{fig-ext:Manipulation_zone}
\end{figure*}

\clearpage

\makeatletter
\renewcommand \thesubsection{S\@arabic\c@subsection}
\setcounter{subsection}{0}
\makeatother

\section*{Supplementary information}
\subsection{Dealing with random valley splitting due to alloy disorder}
\label{sec:sup_valley}
An important aspect for the realization of any shuttling-based architecture in the Si/SiGe platform will likely be the distribution of the variation of the valley splitting in Si/SiGe heterostructures. A recent publication by Wuetz \textit{et al.} \cite{Wuetz2022} suggests that the valley splitting is dominated by alloy disorder and thus varies randomly across different quantum dot positions. Occasional small valley splittings are unfavorable for resonant EDSR-driving with fast electron movement and shuttling coherence, but can be mitigated by adjusting the shuttling path and manipulation position. Possible modifications of the heterostructure, including Ge spikes\cite{McJunkin2021}, increased\cite{Wuetz2022} or oscillating\cite{Mcjunkin2022} Ge concentrations inside the quantum well, may reduce the occurrence of small valley splittings or ideally avoid them altogether. Analogously, the effect of occasional detrimental defects located in the shuttling channel can as well be mitigated by optimizing the shuttling path. In the worst case, a faulty unit cell can be omitted by mapping an algorithm to physical qubits differently.

\subsection{Wiring complexity and solutions}
\label{sec:sup_wiring}
As the wiring density can be a main limiting factor for the size of processor achievable with a given integration approach, we present an estimate of the number of required signals based on an economical operating strategy. The shuttling lanes in Fig. \ref{fig:QuBus_Arch}c and d are color-coded accordingly to indicate which shuttling lanes are controlled  globally or at the individual unit cell level. The vertical shuttling lanes colored green are intended for the connection of unit cells and are controlled at a global level, so that four shared AC signals can suffice. The envisioned operating mode is that these shared lanes periodically move up and down by one or several unit cell sizes and that electrons can be transferred onto them from the horizontal red and blue lanes, which are controlled individually for each unit cell. Either of these individually controlled shuttling lanes is also used for single-qubit gates, while both of them are utilized for a two-qubit gate at their junction. The use of the same signals for the vertical and horizontal red sections in one unit cell is enabled by the assumption that the IR zone is only used when no qubit in the horizontal red lane and no qubit in the blue lane is near the T-junction to the IR zone when the red horizontal lane is active. For the independent operation of the IR zones and to compensate disorder, we also propose local control of the first two individual clavier gates in the shuttling lane and of the SET plunger gate to maintain a sensitive operating point. In addition, one readout wire per cell needs to have a Megahertz-scale bandwidth. If the shuttling path needs to be controlled to avoid disorder or sites with low valley splitting (section \ref{sec:sup_valley}), a few signals per unit cell applied to the screening gates need to be added, with the exact number depending on the required degree of independent control. We thus estimate that a quantum processor chip with $N$ unit cells requires $14\,N+4$ AC signals. 

Analogously, DC voltages are also either shared across all unit cells or individually provided for each unit cell. For the screening gates defining the shuttling channels, inducing gates for SET reservoirs and the global top gate, three global DC voltages are sufficient. For adjusting the reservoir potential to the shuttling lanes, the voltage on the second Ohmic contact to the SET should be tunable. In addition, two SET barrier gates must be tuned. T-junctions and manipulation zones do not require additional DC voltages. Lastly, grounding of the micromagnets is optional and can be shared across all micromagnets. Hence, $3\,N+4$ DC voltages are needed.

While there are no inherent scaling limitations beyond gate and shuttling errors to our architecture at the quantum layer, the wiring requirements have to be compatible with cryostat wiring, packaging and back-end-of-line (BEOL) technology. We estimate that currently available wiring solutions in cryostats of about 1,000 coaxial cables\cite{Bluefors} are the main limiting factor and can accommodate a quantum processor chip with $9 \times 8 = 72$ unit cells. The highest on-chip wiring density occurs at the edge of the quantum plane and requires a few wiring layers for fan-out, which can be realized using standard BEOL technology. Assuming a unit cell size of $\SI{10}{\micro\meter} \times \SI{10}{\micro \meter}$ and a conservative wiring pitch of $\SI{500}{\nano\meter}$ in the on-chip wiring layers, each wiring layer can accommodate 720 lanes along the circumference of the quantum layer, whereas unit cells provide sufficient space to extend all vias across the required number of wiring layers. Thus, a few wiring layers are sufficient to route all signals to the edges of the quantum processor chip or spread them across the surface depending on the connection scheme.

For connecting the chip to a printed circuit board (PCB), high-density socket solutions are an established technology enabling as high as 6,096 connections in consumer devices\cite{AMDSP5}. Commercial land grid array (LGA) solutions that allow for chip-to-board connection are available with contact pitches as low as \SI{0.4}{\mm} and are validated for cryogenic operation\cite{Ardentconcepts}. Assuming a dense connection arrangement and neglecting additional ground connections this results in a chip size of \SI{14}{\mm} by \SI{14}{\mm}. Finally, we estimate the footprint of the connection from the PCB to the coaxial wiring in the cryostat based on current sub-miniature coaxial connectors. 
For example, Rosenberger WSMP\textregistered \ connectors can feature a center-to-center spacing of $\SI{2.15}{\milli \meter}$ while maintaining the signal integrity of a coaxial connector for frequencies up to $\SI{100}{\giga \Hz}$ \cite{Rosenberger}. This results in a PCB area of $\sim \SI{50}{\cm \squared}$ for 1000 connectors, including some overhead for screws etc. Flex-cable solutions can enable an even higher density \cite{Cummings2022}. DC wiring using flex-connectors will have a negligible footprint compared to the high frequency connections.

\subsection{Scaling perspective using cryoelectronic control circuits}
\label{sec:sup_scale}
The proposed layout is well suited for implementing surface codes for quantum error correction, and the predicted fidelities are in the range of what is needed to achieve a reasonable logical qubit performance and overhead. Individual logical qubits are thus within reach for qubit numbers that are reachable with conventional control and packaging and approaches as outlined above. For large scale quantum computing with many error corrected qubits, however, such solutions are less appealing. Here, integrated control solutions offer an attractive pathway. The SpinBus architecture features good prospects in this respect because the purely capacitive impedance of control electrodes, low operating frequency and robust coherence of spin qubits facilitate the use of cryogenic CMOS control circuits. The variable unit cell size can be adjusted to the required size of dedicated control circuits for each unit cell, so that direct wiring, e.g., via flip chip bonding, can eliminate the wiring fan out problem. First estimates of the size of control circuits for spin qubits lead to values in a compatible range \cite{Geck2019}. Next to the size of control circuits, their power dissipation will be a concern in the light of limited cooling power at low temperature. For DC bias, a consumption at level of a few $\SI{}{\nano\watt}$ per channel has already been shown \cite{Otten2022}. Furthermore, thermally isolating flip-chip solutions may allow the operation of electronics at a higher temperature than the qubits \cite{Zoschke2019, Bluhm2021}, making cooling powers potentially at the levels of Watts accessible. While it remains to be seen if the dynamic qubit and shuttling control signals can be generated within the resulting power budget, a possible approach is to select from externally generated pulses using simple cryo-CMOS switches and tuning the qubit response to these fixed pulses using DC gate voltages \cite{Pauka2021, Bohuslavskyi2022}. For readout, heterojunction-bipolar-transistors (HBTs) allow single-shot readout in less than \SI{10}{\micro \s} at powers below \SI{800}{\nano \W} \cite{Curry2019}, which likely can be reduced with optimized sensor designs \cite{Kammerloher2023, Seidler2023}.


\begin{thebibliography}{68}
\makeatletter
\providecommand \@ifxundefined [1]{%
 \@ifx{#1\undefined}
}%
\providecommand \@ifnum [1]{%
 \ifnum #1\expandafter \@firstoftwo
 \else \expandafter \@secondoftwo
 \fi
}%
\providecommand \@ifx [1]{%
 \ifx #1\expandafter \@firstoftwo
 \else \expandafter \@secondoftwo
 \fi
}%
\providecommand \natexlab [1]{#1}%
\providecommand \enquote  [1]{``#1''}%
\providecommand \bibnamefont  [1]{#1}%
\providecommand \bibfnamefont [1]{#1}%
\providecommand \citenamefont [1]{#1}%
\providecommand \href@noop [0]{\@secondoftwo}%
\providecommand \href [0]{\begingroup \@sanitize@url \@href}%
\providecommand \@href[1]{\@@startlink{#1}\@@href}%
\providecommand \@@href[1]{\endgroup#1\@@endlink}%
\providecommand \@sanitize@url [0]{\catcode `\\12\catcode `\$12\catcode
  `\&12\catcode `\#12\catcode `\^12\catcode `\_12\catcode `\%12\relax}%
\providecommand \@@startlink[1]{}%
\providecommand \@@endlink[0]{}%
\providecommand \url  [0]{\begingroup\@sanitize@url \@url }%
\providecommand \@url [1]{\endgroup\@href {#1}{\urlprefix }}%
\providecommand \urlprefix  [0]{URL }%
\providecommand \Eprint [0]{\href }%
\providecommand \doibase [0]{https://doi.org/}%
\providecommand \selectlanguage [0]{\@gobble}%
\providecommand \bibinfo  [0]{\@secondoftwo}%
\providecommand \bibfield  [0]{\@secondoftwo}%
\providecommand \translation [1]{[#1]}%
\providecommand \BibitemOpen [0]{}%
\providecommand \bibitemStop [0]{}%
\providecommand \bibitemNoStop [0]{.\EOS\space}%
\providecommand \EOS [0]{\spacefactor3000\relax}%
\providecommand \BibitemShut  [1]{\csname bibitem#1\endcsname}%
\let\auto@bib@innerbib\@empty

\bibitem{Campbell2017}
\bibinfo{author}{Campbell, E.~T.}, \bibinfo{author}{Terhal, B.~M.} \&
  \bibinfo{author}{Vuillot, C.}
\newblock \bibinfo{title}{Roads towards fault-tolerant universal quantum
  computation}.
\newblock \emph{\bibinfo{journal}{Nature}} \textbf{\bibinfo{volume}{549}},
  \bibinfo{pages}{172--179} (\bibinfo{year}{2017}).

\bibitem{Zwerver2022}
\bibinfo{author}{Zwerver, A. M.~J.} \emph{et~al.}
\newblock \bibinfo{title}{Qubits made by advanced semiconductor manufacturing}.
\newblock \emph{\bibinfo{journal}{Nat. Electron.}}
  \textbf{\bibinfo{volume}{5}}, \bibinfo{pages}{184--190}
  (\bibinfo{year}{2022}).

\bibitem{Yoneda2018}
\bibinfo{author}{Yoneda, J.} \emph{et~al.}
\newblock \bibinfo{title}{A quantum-dot spin qubit with coherence limited by
  charge noise and fidelity higher than 99.9\%}.
\newblock \emph{\bibinfo{journal}{Nat. Nanotechnol.}}
  \textbf{\bibinfo{volume}{13}}, \bibinfo{pages}{102--106}
  (\bibinfo{year}{2018}).

\bibitem{Zajac2018}
\bibinfo{author}{Zajac, D.~M.} \emph{et~al.}
\newblock \bibinfo{title}{Resonantly driven {CNOT} gate for electron spins}.
\newblock \emph{\bibinfo{journal}{Science}} \textbf{\bibinfo{volume}{359}},
  \bibinfo{pages}{439--442} (\bibinfo{year}{2018}).

\bibitem{Watson2018}
\bibinfo{author}{Watson, T.~F.} \emph{et~al.}
\newblock \bibinfo{title}{A programmable two-qubit quantum processor in
  silicon}.
\newblock \emph{\bibinfo{journal}{Nature}} \textbf{\bibinfo{volume}{555}},
  \bibinfo{pages}{633--637} (\bibinfo{year}{2018}).

\bibitem{Xue2019}
\bibinfo{author}{Xue, X.} \emph{et~al.}
\newblock \bibinfo{title}{Benchmarking gate fidelities in a
  $\mathrm{Si}/\mathrm{SiGe}$ two-qubit device}.
\newblock \emph{\bibinfo{journal}{Phys. Rev. X}} \textbf{\bibinfo{volume}{9}},
  \bibinfo{pages}{021011} (\bibinfo{year}{2019}).

\bibitem{Petit2020}
\bibinfo{author}{Petit, L.} \emph{et~al.}
\newblock \bibinfo{title}{Universal quantum logic in hot silicon qubits}.
\newblock \emph{\bibinfo{journal}{Nature}} \textbf{\bibinfo{volume}{580}},
  \bibinfo{pages}{355--359} (\bibinfo{year}{2020}).

\bibitem{Xue2022}
\bibinfo{author}{Xue, X.} \emph{et~al.}
\newblock \bibinfo{title}{Quantum logic with spin qubits crossing the surface
  code threshold}.
\newblock \emph{\bibinfo{journal}{Nature}} \textbf{\bibinfo{volume}{601}},
  \bibinfo{pages}{343--347} (\bibinfo{year}{2022}).

\bibitem{Noiri2022-1}
\bibinfo{author}{Noiri, A.} \emph{et~al.}
\newblock \bibinfo{title}{Fast universal quantum gate above the fault-tolerance
  threshold in silicon}.
\newblock \emph{\bibinfo{journal}{Nature}} \textbf{\bibinfo{volume}{601}},
  \bibinfo{pages}{338–342} (\bibinfo{year}{2022}).

\bibitem{Takeda2022}
\bibinfo{author}{Takeda, K.}, \bibinfo{author}{Noiri, A.},
  \bibinfo{author}{Nakajima, T.}, \bibinfo{author}{Kobayashi, T.} \&
  \bibinfo{author}{Tarucha, S.}
\newblock \bibinfo{title}{Quantum error correction with silicon spin qubits}.
\newblock \emph{\bibinfo{journal}{Nature}} \textbf{\bibinfo{volume}{608}},
  \bibinfo{pages}{682--686} (\bibinfo{year}{2022}).

\bibitem{Mills2022}
\bibinfo{author}{Mills, A.~R.} \emph{et~al.}
\newblock \bibinfo{title}{Two-qubit silicon quantum processor with operation
  fidelity exceeding 99\%}.
\newblock \emph{\bibinfo{journal}{Sci. Adv.}} \textbf{\bibinfo{volume}{8}},
  \bibinfo{pages}{eabn5130} (\bibinfo{year}{2022}).

\bibitem{Lawrie2020}
\bibinfo{author}{Lawrie, W. I.~L.} \emph{et~al.}
\newblock \bibinfo{title}{Quantum dot arrays in silicon and germanium}.
\newblock \emph{\bibinfo{journal}{Appl. Phys. Lett.}}
  \textbf{\bibinfo{volume}{116}}, \bibinfo{pages}{080501}
  (\bibinfo{year}{2020}).

\bibitem{Mortemousque2021}
\bibinfo{author}{Mortemousque, P.-A.} \emph{et~al.}
\newblock \bibinfo{title}{Coherent control of individual electron spins in a
  two-dimensional array of quantum dots}.
\newblock \emph{\bibinfo{journal}{Nat. Nanotechnol.}}
  \textbf{\bibinfo{volume}{16}}, \bibinfo{pages}{296–301}
  (\bibinfo{year}{2021}).

\bibitem{Weinstein2023}
\bibinfo{author}{Weinstein, A.~J.} \emph{et~al.}
\newblock \bibinfo{title}{Universal logic with encoded spin qubits in silicon}.
\newblock \emph{\bibinfo{journal}{Nature}} \textbf{\bibinfo{volume}{615}},
  \bibinfo{pages}{817--822} (\bibinfo{year}{2023}).

\bibitem{Philips2022}
\bibinfo{author}{Philips, S. G.~J.} \emph{et~al.}
\newblock \bibinfo{title}{Universal control of a six-qubit quantum processor in
  silicon}.
\newblock \emph{\bibinfo{journal}{Nature}} \textbf{\bibinfo{volume}{609}},
  \bibinfo{pages}{919--924} (\bibinfo{year}{2022}).

\bibitem{Undseth2023}
\bibinfo{author}{Undseth, B.} \emph{et~al.}
\newblock \bibinfo{title}{Nonlinear response and crosstalk of electrically
  driven silicon spin qubits}.
\newblock \emph{\bibinfo{journal}{Phys. Rev. Appl.}}
  \textbf{\bibinfo{volume}{19}}, \bibinfo{pages}{044078}
  (\bibinfo{year}{2023}).

\bibitem{Vandersypen2017}
\bibinfo{author}{Vandersypen, L.} \emph{et~al.}
\newblock \bibinfo{title}{Interfacing spin qubits in quantum dots and
  donors—hot, dense, and coherent}.
\newblock \emph{\bibinfo{journal}{npj Quantum Inf.}}
  \textbf{\bibinfo{volume}{3}}, \bibinfo{pages}{34} (\bibinfo{year}{2017}).

\bibitem{Li2018}
\bibinfo{author}{Li, R.} \emph{et~al.}
\newblock \bibinfo{title}{A crossbar network for silicon quantum dot qubits}.
\newblock \emph{\bibinfo{journal}{Sci. Adv.}} \textbf{\bibinfo{volume}{4}},
  \bibinfo{pages}{eaar3960} (\bibinfo{year}{2018}).

\bibitem{Veldhorst2017}
\bibinfo{author}{Veldhorst, M.}, \bibinfo{author}{Eenink, H.},
  \bibinfo{author}{Yang, C.} \& \bibinfo{author}{Dzurak, A.}
\newblock \bibinfo{title}{Silicon cmos architecture for a spin-based quantum
  computer}.
\newblock \emph{\bibinfo{journal}{Nat. Commun.}} \textbf{\bibinfo{volume}{8}},
  \bibinfo{pages}{1766} (\bibinfo{year}{2017}).

\bibitem{Geck2019}
\bibinfo{author}{Geck, L.}, \bibinfo{author}{Kruth, A.},
  \bibinfo{author}{Bluhm, H.}, \bibinfo{author}{van Waasen, S.} \&
  \bibinfo{author}{Heinen, S.}
\newblock \bibinfo{title}{Control electronics for semiconductor spin qubits}.
\newblock \emph{\bibinfo{journal}{Quantum Sci. Technol.}}
  \textbf{\bibinfo{volume}{5}}, \bibinfo{pages}{015004} (\bibinfo{year}{2020}).

\bibitem{Boter2022}
\bibinfo{author}{Boter, J.~M.} \emph{et~al.}
\newblock \bibinfo{title}{Spiderweb array: A sparse spin-qubit array}.
\newblock \emph{\bibinfo{journal}{Phys. Rev. Appl.}}
  \textbf{\bibinfo{volume}{18}}, \bibinfo{pages}{024053}
  (\bibinfo{year}{2022}).

\bibitem{Baart2016}
\bibinfo{author}{Baart, T.~A.} \emph{et~al.}
\newblock \bibinfo{title}{Single-spin {CCD}}.
\newblock \emph{\bibinfo{journal}{Nat. Nanotechnol.}}
  \textbf{\bibinfo{volume}{11}}, \bibinfo{pages}{330--334}
  (\bibinfo{year}{2016}).

\bibitem{Flentje2017}
\bibinfo{author}{Flentje, H.} \emph{et~al.}
\newblock \bibinfo{title}{Coherent long-distance displacement of individual
  electron spins}.
\newblock \emph{\bibinfo{journal}{Nat. Commun.}} \textbf{\bibinfo{volume}{8}},
  \bibinfo{pages}{501} (\bibinfo{year}{2017}).

\bibitem{Mills2019}
\bibinfo{author}{Mills, A.} \emph{et~al.}
\newblock \bibinfo{title}{Shuttling a single charge across a one-dimensional
  array of silicon quantum dots}.
\newblock \emph{\bibinfo{journal}{Nat Commun.}} \textbf{\bibinfo{volume}{10}},
  \bibinfo{pages}{1063} (\bibinfo{year}{2019}).

\bibitem{Yoneda2021}
\bibinfo{author}{Yoneda, J.} \emph{et~al.}
\newblock \bibinfo{title}{Coherent spin qubit transport in silicon}.
\newblock \emph{\bibinfo{journal}{Nat. Commun.}} \textbf{\bibinfo{volume}{12}},
  \bibinfo{pages}{4114} (\bibinfo{year}{2021}).

\bibitem{Noiri2022-2}
\bibinfo{author}{Noiri, A.} \emph{et~al.}
\newblock \bibinfo{title}{A shuttling-based two-qubit logic gate for linking
  distant silicon quantum processors}.
\newblock \emph{\bibinfo{journal}{Nat. Commun.}} \textbf{\bibinfo{volume}{13}},
  \bibinfo{pages}{5740} (\bibinfo{year}{2022}).

\bibitem{Zwerver2022-2}
\bibinfo{author}{Zwerver, A.} \emph{et~al.}
\newblock \bibinfo{title}{Shuttling an electron spin through a silicon quantum
  dot array}.
\newblock \bibinfo{howpublished}{Preprint at https://arxiv.org/abs/2209.00920}
  (\bibinfo{year}{2022}).

\bibitem{Seidler2021}
\bibinfo{author}{Seidler, I.} \emph{et~al.}
\newblock \bibinfo{title}{Conveyor-mode single-electron shuttling in
  $\mathrm{Si}/\mathrm{SiGe}$ for a scalable quantum computing architecture}.
\newblock \emph{\bibinfo{journal}{npj Quantum Inf.}}
  \textbf{\bibinfo{volume}{8}}, \bibinfo{pages}{100} (\bibinfo{year}{2022}).

\bibitem{Xue2023}
\bibinfo{author}{Xue, R.} \emph{et~al.}
\newblock \bibinfo{title}{$\mathrm{Si}/\mathrm{SiGe}$ {{QuBus}} with memory and
  connectivity function for single carrier information processing devices over
  micron scale distances}.
\newblock \bibinfo{howpublished}{In preparation} (\bibinfo{year}{2023}).

\bibitem{Langrock2023}
\bibinfo{author}{Langrock, V.} \emph{et~al.}
\newblock \bibinfo{title}{Blueprint of a scalable spin qubit shuttle device for
  coherent mid-range qubit transfer in disordered
  $\mathrm{Si}/\mathrm{SiGe}/\mathrm{SiO_2}$}.
\newblock \emph{\bibinfo{journal}{PRX Quantum}} \textbf{\bibinfo{volume}{4}},
  \bibinfo{pages}{020305} (\bibinfo{year}{2023}).

\bibitem{Struck2023}
\bibinfo{author}{Struck, T.} \emph{et~al.}
\newblock \bibinfo{title}{Spin-$\mathrm{EPR}$-pair separation by conveyor-mode
  single electron shuttling in $\mathrm{Si}/\mathrm{SiGe}$}.
\newblock \bibinfo{howpublished}{In preparation} (\bibinfo{year}{2023}).

\bibitem{Ono2002}
\bibinfo{author}{Ono, K.}, \bibinfo{author}{Austing, D.~G.},
  \bibinfo{author}{Tokura, Y.} \& \bibinfo{author}{Tarucha, S.}
\newblock \bibinfo{title}{Current rectification by pauli exclusion in a weakly
  coupled double quantum dot system}.
\newblock \emph{\bibinfo{journal}{Science}} \textbf{\bibinfo{volume}{297}},
  \bibinfo{pages}{1313--1317} (\bibinfo{year}{2002}).

\bibitem{Connors2020}
\bibinfo{author}{Connors, E.~J.}, \bibinfo{author}{Nelson, J.} \&
  \bibinfo{author}{Nichol, J.~M.}
\newblock \bibinfo{title}{Rapid high-fidelity spin-state readout in
  $\mathrm{Si}$/$\mathrm{Si}$-$\mathrm{Ge}$ quantum dots via rf reflectometry}.
\newblock \emph{\bibinfo{journal}{Phys. Rev. Appl.}}
  \textbf{\bibinfo{volume}{13}}, \bibinfo{pages}{024019}
  (\bibinfo{year}{2020}).

\bibitem{Vink2007}
\bibinfo{author}{Vink, I.~T.}, \bibinfo{author}{Nooitgedagt, T.},
  \bibinfo{author}{Schouten, R.~N.}, \bibinfo{author}{Vandersypen, L. M.~K.} \&
  \bibinfo{author}{Wegscheider, W.}
\newblock \bibinfo{title}{Cryogenic amplifier for fast real-time detection of
  single-electron tunneling}.
\newblock \emph{\bibinfo{journal}{Appl. Phys. Lett.}}
  \textbf{\bibinfo{volume}{91}}, \bibinfo{pages}{123512}
  (\bibinfo{year}{2007}).

\bibitem{Tracy2016}
\bibinfo{author}{Tracy, L.~A.} \emph{et~al.}
\newblock \bibinfo{title}{Single shot spin readout using a cryogenic
  high-electron-mobility transistor amplifier at sub-{{Kelvin}} temperatures}.
\newblock \emph{\bibinfo{journal}{Appl. Phys. Lett.}}
  \textbf{\bibinfo{volume}{108}}, \bibinfo{pages}{063101}
  (\bibinfo{year}{2016}).

\bibitem{Curry2019}
\bibinfo{author}{Curry, M.~J.} \emph{et~al.}
\newblock \bibinfo{title}{Single-{{Shot Readout Performance}} of {{Two
  Heterojunction-Bipolar-Transistor Amplification Circuits}} at {{Millikelvin
  Temperatures}}}.
\newblock \emph{\bibinfo{journal}{Sci. Rep.}} \textbf{\bibinfo{volume}{9}},
  \bibinfo{pages}{16976} (\bibinfo{year}{2019}).

\bibitem{Loss1998}
\bibinfo{author}{Loss, D.} \& \bibinfo{author}{DiVincenzo, D.~P.}
\newblock \bibinfo{title}{Quantum computation with quantum dots}.
\newblock \emph{\bibinfo{journal}{Phys. Rev. A}} \textbf{\bibinfo{volume}{57}},
  \bibinfo{pages}{120--126} (\bibinfo{year}{1998}).

\bibitem{Meunier2011}
\bibinfo{author}{Meunier, T.}, \bibinfo{author}{Calado, V.~E.} \&
  \bibinfo{author}{Vandersypen, L. M.~K.}
\newblock \bibinfo{title}{Efficient controlled-phase gate for single-spin
  qubits in quantum dots}.
\newblock \emph{\bibinfo{journal}{Phys. Rev. B}} \textbf{\bibinfo{volume}{83}},
  \bibinfo{pages}{121403} (\bibinfo{year}{2011}).

\bibitem{Russ2018}
\bibinfo{author}{Russ, M.} \emph{et~al.}
\newblock \bibinfo{title}{High-fidelity quantum gates in
  $\mathrm{Si}/\mathrm{SiGe}$ double quantum dots}.
\newblock \emph{\bibinfo{journal}{Phys. Rev. B}} \textbf{\bibinfo{volume}{97}},
  \bibinfo{pages}{085421} (\bibinfo{year}{2018}).

\bibitem{Petit2022}
\bibinfo{author}{Petit, L.} \emph{et~al.}
\newblock \bibinfo{title}{Design and integration of single-qubit rotations and
  two-qubit gates in silicon above one {{Kelvin}}}.
\newblock \emph{\bibinfo{journal}{Commun. Mater.}}
  \textbf{\bibinfo{volume}{3}}, \bibinfo{pages}{82} (\bibinfo{year}{2022}).

\bibitem{Teske2022}
\bibinfo{author}{Teske, J.~D.}, \bibinfo{author}{Cerfontaine, P.} \&
  \bibinfo{author}{Bluhm, H.}
\newblock \bibinfo{title}{qopt: An experiment-oriented software package for
  qubit simulation and quantum optimal control}.
\newblock \emph{\bibinfo{journal}{Phys. Rev. Applied}}
  \textbf{\bibinfo{volume}{17}}, \bibinfo{pages}{034036}
  (\bibinfo{year}{2022}).

\bibitem{Dial2013}
\bibinfo{author}{Dial, O.~E.} \emph{et~al.}
\newblock \bibinfo{title}{Charge noise spectroscopy using coherent exchange
  oscillations in a singlet-triplet qubit}.
\newblock \emph{\bibinfo{journal}{Phys. Rev. Lett.}}
  \textbf{\bibinfo{volume}{110}} (\bibinfo{year}{2013}).

\bibitem{Kranz2020}
\bibinfo{author}{Kranz, L.} \emph{et~al.}
\newblock \bibinfo{title}{Exploiting a single-crystal environment to minimize
  the charge noise on qubits in silicon}.
\newblock \emph{\bibinfo{journal}{Adv. Mater.}} \textbf{\bibinfo{volume}{32}},
  \bibinfo{pages}{2003361} (\bibinfo{year}{2020}).

\bibitem{Cerfontaine2020}
\bibinfo{author}{Cerfontaine, P.} \emph{et~al.}
\newblock \bibinfo{title}{Closed-loop control of a $\mathrm{GaAs}$-based
  singlet-triplet spin qubit with 99.5 \% gate fidelity and low leakage}.
\newblock \emph{\bibinfo{journal}{Nat. Commun.}} \textbf{\bibinfo{volume}{11}},
  \bibinfo{pages}{4144} (\bibinfo{year}{2022}).

\bibitem{Fowler2012}
\bibinfo{author}{Fowler, A.~G.}, \bibinfo{author}{Mariantoni, M.},
  \bibinfo{author}{Martinis, J.~M.} \& \bibinfo{author}{Cleland, A.~N.}
\newblock \bibinfo{title}{Surface codes: Towards practical large-scale quantum
  computation}.
\newblock \emph{\bibinfo{journal}{Phys. Rev. A}} \textbf{\bibinfo{volume}{86}},
  \bibinfo{pages}{032324} (\bibinfo{year}{2012}).

\bibitem{Bluefors}
\bibinfo{title}{{Bluefors XLDsl Dilution Refrigerator System}}.
\newblock
  \bibinfo{howpublished}{\url{https://bluefors.com/products/xldsl-dilution-refrigerator}}
  (\bibinfo{year}{Accessed: 2023-06-27}).

\bibitem{Donahue1999}
\bibinfo{author}{Donahue, M.}
\newblock \bibinfo{title}{Oommf user's guide, version 1.0}
  (\bibinfo{year}{1999}).

\bibitem{Neumann2015}
\bibinfo{author}{Neumann, R.} \& \bibinfo{author}{Schreiber, L.~R.}
\newblock \bibinfo{title}{Simulation of micro-magnet stray-field dynamics for
  spin qubit manipulation}.
\newblock \emph{\bibinfo{journal}{J. Appl. Phys.}}
  \textbf{\bibinfo{volume}{117}}, \bibinfo{pages}{193903}
  (\bibinfo{year}{2015}).

\bibitem{Friesen2007}
\bibinfo{author}{Friesen, M.}, \bibinfo{author}{Chutia, S.},
  \bibinfo{author}{Tahan, C.} \& \bibinfo{author}{Coppersmith, S.~N.}
\newblock \bibinfo{title}{Valley splitting theory of
  $\mathrm{Si}\mathrm{Ge}/\mathrm{Si}/\mathrm{Si}\mathrm{Ge}$ quantum wells}.
\newblock \emph{\bibinfo{journal}{Phys. Rev. B}} \textbf{\bibinfo{volume}{75}},
  \bibinfo{pages}{115318} (\bibinfo{year}{2007}).

\bibitem{Wood2018}
\bibinfo{author}{Wood, C.~J.} \& \bibinfo{author}{Gambetta, J.~M.}
\newblock \bibinfo{title}{Quantification and characterization of leakage
  errors}.
\newblock \emph{\bibinfo{journal}{Phys. Rev. A}} \textbf{\bibinfo{volume}{97}}
  (\bibinfo{year}{2018}).

\bibitem{Wuetz2022}
\bibinfo{author}{Wuetz, B.~P.} \emph{et~al.}
\newblock \bibinfo{title}{Atomic fluctuations lifting the energy degeneracy in
  $\mathrm{Si}/\mathrm{Si}\mathrm{Ge}$ quantum dots}.
\newblock \emph{\bibinfo{journal}{Nat. Commun.}} \textbf{\bibinfo{volume}{13}},
  \bibinfo{pages}{7730} (\bibinfo{year}{2022}).

\bibitem{Hollmann2020}
\bibinfo{author}{Hollmann, A.} \emph{et~al.}
\newblock \bibinfo{title}{Large, tunable valley splitting and single-spin
  relaxation mechanisms in a
  $\mathrm{Si}$/$\mathrm{Si}_{x}$$\mathrm{Ge}_{1\ensuremath{-}x}$ quantum dot}.
\newblock \emph{\bibinfo{journal}{Phys. Rev. Appl.}}
  \textbf{\bibinfo{volume}{13}}, \bibinfo{pages}{034068}
  (\bibinfo{year}{2020}).

\bibitem{Watts2015}
\bibinfo{author}{Watts, P.} \emph{et~al.}
\newblock \bibinfo{title}{{Optimizing for an arbitrary perfect entangler. I.
  Functionals}}.
\newblock \emph{\bibinfo{journal}{Phys. Rev. A}} \textbf{\bibinfo{volume}{91}}
  (\bibinfo{year}{2015}).

\bibitem{Vandersypen2001}
\bibinfo{author}{Vandersypen, L. M.~K.} \emph{et~al.}
\newblock \bibinfo{title}{Experimental realization of {{Shor's}} quantum
  factoring algorithm using nuclear magnetic resonance}.
\newblock \emph{\bibinfo{journal}{Nature}} \textbf{\bibinfo{volume}{414}},
  \bibinfo{pages}{883--887} (\bibinfo{year}{2001}).

\bibitem{McJunkin2021}
\bibinfo{author}{McJunkin, T.} \emph{et~al.}
\newblock \bibinfo{title}{Valley splittings in $\mathrm{Si}/\mathrm{SiGe}$
  quantum dots with a germanium spike in the silicon well}.
\newblock \emph{\bibinfo{journal}{Phys. Rev. B}}
  \textbf{\bibinfo{volume}{104}}, \bibinfo{pages}{085406}
  (\bibinfo{year}{2021}).

\bibitem{Mcjunkin2022}
\bibinfo{author}{McJunkin, T.} \emph{et~al.}
\newblock \bibinfo{title}{$\mathrm{SiGe}$ quantum wells with oscillating
  $\mathrm{Ge}$ concentrations for quantum dot qubits}.
\newblock \emph{\bibinfo{journal}{Nat. Commun.}} \textbf{\bibinfo{volume}{13}},
  \bibinfo{pages}{1--7} (\bibinfo{year}{2022}).

\bibitem{AMDSP5}
\bibinfo{title}{{AMD EPYC}\texttrademark{} 9654}.
\newblock
  \bibinfo{howpublished}{\url{https://www.amd.com/en/products/cpu/amd-epyc-9654}}
  (\bibinfo{year}{Accessed: 2023-06-27}).

\bibitem{Ardentconcepts}
\bibinfo{title}{{Ardent Concepts High Density TR Multicoax}\texttrademark{}
  {Cabling}}.
\newblock
  \bibinfo{howpublished}{\url{https://www.ardentconcepts.com/quantum-overview}}
  (\bibinfo{year}{Accessed: 2023-06-27}).

\bibitem{Rosenberger}
\bibinfo{title}{{{R}}osenberger {{WSMP}}\textregistered{} {{NEW}}
  {{GENERATION}} connectors}.
\newblock
  \bibinfo{howpublished}{\url{https://www.rosenberger.com/product/wsmp/}}
  (\bibinfo{year}{Accessed: 2023-06-27}).

\bibitem{Cummings2022}
\bibinfo{author}{Cummings, J.~D.}, \bibinfo{author}{Rokosz, J.~A.},
  \bibinfo{author}{Thompson, K.~J.} \& \bibinfo{author}{Weber, S.~J.}
\newblock \bibinfo{title}{High-{{Density Cryogenic Wiring}} for
  {{Superconducting Qubit Control}}}.
\newblock \bibinfo{howpublished}{Patent US20220230785A1}
  (\bibinfo{year}{2022}).

\bibitem{Otten2022}
\bibinfo{author}{Otten, R.} \emph{et~al.}
\newblock \bibinfo{title}{Qubit {{Bias}} using a {{CMOS DAC}} at {{mK
  Temperatures}}}.
\newblock In \emph{\bibinfo{booktitle}{2022 29th {{IEEE International
  Conference}} on {{Electronics}}, {{Circuits}} and {{Systems}} ({{ICECS}})}}
  (\bibinfo{year}{2022}).

\bibitem{Zoschke2019}
\bibinfo{author}{Zoschke, K.} \emph{et~al.}
\newblock \bibinfo{title}{High {{Density Flex}} and {{Thin Chip Embedding
  Technology}} for {{Polymeric Interposer}} and {{Sensor Packaging
  Applications}}}.
\newblock In \emph{\bibinfo{booktitle}{2019 {{International Wafer Level
  Packaging Conference}} ({{IWLPC}})}}, \bibinfo{pages}{1--9}
  (\bibinfo{year}{2019}).

\bibitem{Bluhm2021}
\bibinfo{author}{Bluhm, H.}
\newblock \bibinfo{title}{{{Isolator}} für kryoelektrische {{Chips}} bei
  extrem niedrigen {{Temperaturen}} unter 10{{K}}}.
\newblock \bibinfo{howpublished}{Patent DE102021123046} (\bibinfo{year}{2021}).

\bibitem{Pauka2021}
\bibinfo{author}{Pauka, S.~J.} \emph{et~al.}
\newblock \bibinfo{title}{A cryogenic {{CMOS}} chip for generating control
  signals for multiple qubits}.
\newblock \emph{\bibinfo{journal}{Nat. Electron.}}
  \textbf{\bibinfo{volume}{4}}, \bibinfo{pages}{64--70} (\bibinfo{year}{2021}).

\bibitem{Bohuslavskyi2022}
\bibinfo{author}{Bohuslavskyi, H.} \emph{et~al.}
\newblock \bibinfo{title}{Scalable on-chip multiplexing of low-noise silicon
  electron and hole quantum dots}.
\newblock \bibinfo{howpublished}{Preprint at https://arxiv.org/abs/2208.12131}
  (\bibinfo{year}{2022}).

\bibitem{Kammerloher2023}
\bibinfo{author}{Kammerloher, E.} \emph{et~al.}
\newblock \bibinfo{title}{Sensing dot with high output swing for scalable
  baseband readout of spin qubits}.
\newblock \bibinfo{howpublished}{Preprint at https://arxiv.org/abs/2107.13598}
  (\bibinfo{year}{2023}).

\bibitem{Seidler2023}
\bibinfo{author}{Seidler, I.} \emph{et~al.}
\newblock \bibinfo{title}{Tailoring potentials by simulation-aided design of
  gate layouts for spin qubit applications}.
\newblock \bibinfo{howpublished}{Preprint at https://arxiv.org/abs/2303.13358}
  (\bibinfo{year}{2023}).

\end{thebibliography}
\end{document}